# Superconducting phase diagram of finite-layer nickelates $Nd_{n+1}Ni_nO_{2n+2}$

Check for updates

**Andreas Hausoel** [1,5], **Simone Di Cataldo** [2,5], **Motoharu Kitatani** [3], **Oleg Janson** [1] & **Karsten Held** [4] ✉

Following the successful prediction of the superconducting phase diagram for infinite-layer nickelates, here we calculate the superconducting $T_c$ vs. the number of layers $n$ for finite-layer nickelates using the dynamical vertex approximation. To this end, we start with density functional theory, and include local correlations non-perturbatively by dynamical mean-field theory for $n = 2$–$7$. For all $n$, the Ni $d_{x^2-y^2}$ orbital crosses the Fermi level, but for $n > 4$ there are additional $(\pi, \pi)$ pockets or tubes that slightly enhance the layer-averaged hole doping of the $d_{x^2-y^2}$ orbitals beyond the leading $1/n$ contribution stemming from the valence electron count. We finally calculate $T_c$ for the single-orbital $d_{x^2-y^2}$ Hubbard model by dynamical vertex approximation.

The discovery of superconductivity in nickelates by Li, Hwang et al.[1] has opened a new and very active area of research. In fact, superconducting nickelates remain a largely unexplored family of strongly-correlated systems, and offer researchers a unique and novel testbed to study unconventional superconductivity.

However, synthesizing infinite-layer nickelates $(Ca,Sr)_xR_{1-x}NiO_2$ with rare earth $R$ = Nd, La, Pr, Sm that are actually superconducting has proved to be challenging, and only a few groups succeeded[2–14]. The primary obstacle to synthesize superconducting nickelate films and bulk materials is the large degree of disorder, stacking faults and off-stoichiometry. This is caused by the complicated synthesis process which involves first growing $(Ca,Sr)_xR_{1-x}NiO_3$ and then reducing it to $(Ca,Sr)_xR_{1-x}NiO_2$ with $CaH_2$ or H[13,15]. The necessity to hole-dope the system into the superconducting state with Ca (or Sr) increases the chances of disorder. This, in combination with the oxygen reduction, prevents synthesizing bulk nickelates with $x > 8\%$[16]. It is not surprising that, even among the superconducting films, there is a large variance in the measured properties. Cleaner films show resistivities lower by a factor of three, while at the same time, the critical temperature $T_c$ is considerably larger[9,10].

Against this background, growing finite-layer nickelate films $Nd_{n+1}Ni_nO_{2n+2}$, see Fig. 1 (top), offers a welcome alternative to doping infinite-layer nickelates with Sr or Ca, eliminating at least Sr/Ca disorder. This second family of nickelates presents a series of finite slabs separated by a stacking fault (black dashed lines in Fig. 1) that is delineated by an additional O layer and a shift by half-of-the-unit-cell diagonal perpendicular to the stacking fault. These finite-layer nickelates are a generalization of infinite-layer ones that are recovered for $n \to \infty$. These finite-layer nickelates are not to be confused with the Ruddlesden-Popper series $La_{n+1}Ni_nO_{3n+1}$[17] whose Ni $3d$ shells are filled with $(7 + \frac{1}{n})$ electrons, i.e., between 7 and 8, instead of $(9 - \frac{1}{n})$ for the finite-layer nickelates considered here.

Pan, Mundy et al. established the presence of superconductivity in finite-layer nickelates for $n = 5$[6] and reported ongoing work for other $n$[18]. Similar layered compounds can also form with copper, such as $Tl_2Ba_2Ca_{n-1}Cu_nO_{2n+4}$[19] and $HgBa_2Ca_{n-1}Cu_nO_{2n+2}$[20], in which the $T_c$ is typically maximal for $n = 3$. In a fashion similar to cuprates, it is likely that finite-layer nickelates are far from having reached the optimum $T_c$, as both the number of layers and sample quality can still be optimized.

Moving on to the theoretical studies, superconductivity in bulk nickelates[21] and nickelate heterostructures[22] has been conjectured long before samples were successfully synthesized. Only after their synthesis in 2019[1], however, were they thoroughly examined on the theoretical side, turning them into the *hot* topic they currently are[23–33].

The arguably simplest description for $3d^9$ nickelates is that of a one-band Hubbard model plus largely detached pockets originating from the Nd $5d$ orbitals[34,35]. This simple description was recently confirmed by angular-resolved photoemission spectroscopy (ARPES) that only showed a Ni $3d_{x^2-y^2}$ Fermi surface and a pocket at $(\pi, \pi, \pi)$ ("A pocket") for $Sr_xLa_{1-x}NiO_2$[36,37], in excellent agreement with a priori density functional theory (DFT) plus dynamical mean-field theory (DMFT) calculations[34,38]. Even the superconducting dome determined with the dynamical vertex approximation (D$\Gamma$A)[34] well agrees with *a posteriori* experiments[9], as does the resonant inelastic X-ray spectrum[39,40]. Finite-layer nickelates have been also studied theoretically[41,42], though hitherto with a focus on the pentalayer found in the first experiment[6].

In this letter, we calculated the $T_c$ vs. number of layers $n$ phase diagram for finite-layer neodymium nickelates on a neodymium gallium oxide substrate. We employed DFT to determine the crystal structure of such strained superlattices and calculate their electronic structure. From there, we included local correlations using DMFT in Ni 3$d$ and Nd 5$d$ shells, and finally mapped the result to the minimal single-band model to estimate the superconducting $T_c$ by DΓA. This brings us to the central result of our work—the phase diagram in Fig. 1, showing the dome-like behavior of the superconducting $T_c$ vs. an increasing (from left to right) number of layers $n$. Superconductivity is found for a similar number of layers as in experiment ($n$ = 4–8[18]).

[1]Institute for Theoretical Solid State Physics, Leibniz Institute for Solid State and Materials Research Dresden, Dresden, Germany. [2]Dipartimento di Fisica, Sapienza Università di Roma, Roma, Italy. [3]Department of Material Science, University of Hyogo, Ako, Hyogo, Japan. [4]Institute of Solid State Physics, TU Wien, Vienna, Austria. [5]These authors contributed equally: Andreas Hausoel, Simone Di Cataldo. ✉e-mail: held@ifp.tuwien.ac.at

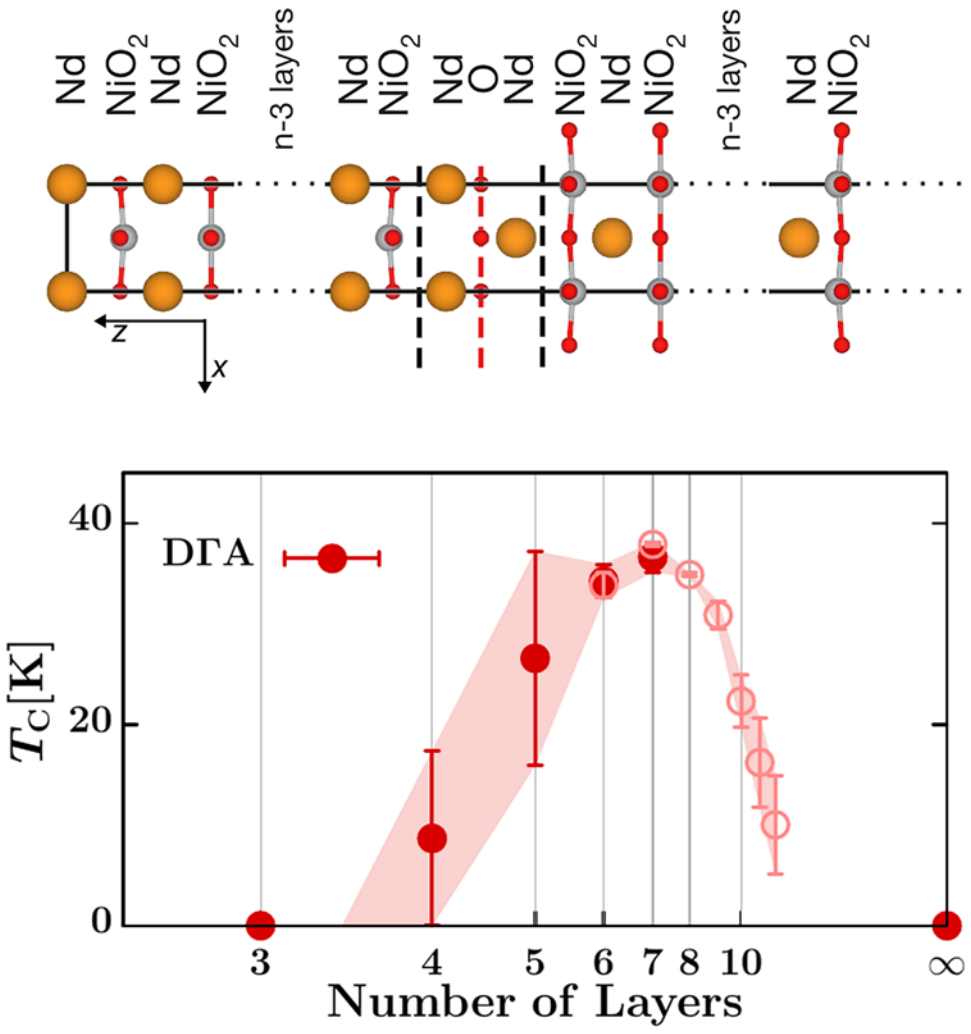

**Fig. 1 | Phase diagram of finite-layer nickelates $Nd_{n+1}Ni_nO_{2n+2}$.** Top: Crystal structure of finite-layer nickelates $Nd_{n+1}Ni_nO_{2n+2}$. Nd, Ni, and O atoms are indicated as large orange, medium gray, and small red spheres, respectively. Dashed black lines indicate the Nd–O–Nd stacking fault. Bottom: Phase diagram superconducting $T_c$ vs. number of layers $n$ as calculated by DΓA. The error bar reflects the uncertainty in the hole doping of the $d_{x^2-y^2}$ orbital, as discussed in Section "Dynamical vertex approximation"; the open circle for $n \geq 6$ indicates that we here used an infinite-layer calculation with total doping $1/n$ to mimic the finite layer nickelate.

## Results

### Density functional theory

The crystal structures for the $Nd_{n+1}Ni_nO_{2n+2}$ series were constructed starting from a supercell of $n$ infinite-layer unit cells and adding the stacking fault. The in-plane lattice parameters were constrained to those of cubic $NdGaO_3$ (NGO), i.e., $a = b = 3.83$ Å, which experimentally is used as a substrate[6,18], while the $c$ axis was relaxed as in ref. 43, see Supplementary Note 2.

There is an out-of-plane distortion of the Ni-O bonds, clearly visible in the true-to-scale Fig. 1 (top), which is particularly strong for the layers facing the stacking fault (4°), and rapidly fades in further layers. See Supplementary Note 2 for details on the distortion angles and further information on the DFT relaxation; and Section IV for the methods and codes used. The interaction with the stacking fault causes a substantial difference between *outer* (close to the stacking fault) and *inner* (away from the stacking fault) Ni-O layers. This interaction also introduces a symmetry inequivalence between odd-$n$ and even-$n$ cases, for which $n = 2$ represents a further special case in which only outer layers are present.

In Fig. 2, we report the DFT-calculated band structure and density of states (DOS), projected onto the most relevant atomic orbitals for $n = 4$ and $n = 7$ layers (for the other values of $n$ we refer the reader to the Supplementary Note 3). For both $n$'s, bands of Ni $d_{x^2-y^2}$ and Nd $d_{xy}$ character cross the Fermi energy. The latter forms the A pocket of the infinite-layer nickelate which becomes one or several M–A tubes for the finite-layer compound, see Supplementary Note 3. Without the stacking fault, we would have just the infinite-layer structure, but with a backfolding of $n$ $k_z$'s into $n$ vineyard-like bands for a supercell of $n$ layers. We see this general pattern, but the stacking fault also leads to marked differences, particularly for small $n$. Most noteworthy, the electron pocket at the $\Gamma$ pocket formed by Nd with some (admixed) Ni $d_{z^2}$ character, which is still present for $n = 7$, gets lifted above the Fermi energy for $n = 4$. The projected DOS in Fig. 2 (left) also shows predominantly Ni $d_{x^2-y^2}$ character at the Fermi level, with Ni $t_{2g}$ and $d_{z^2}$ bands further below and Nd bands above. Changes of the DOS are quite minute when changing the number of layers $n$.

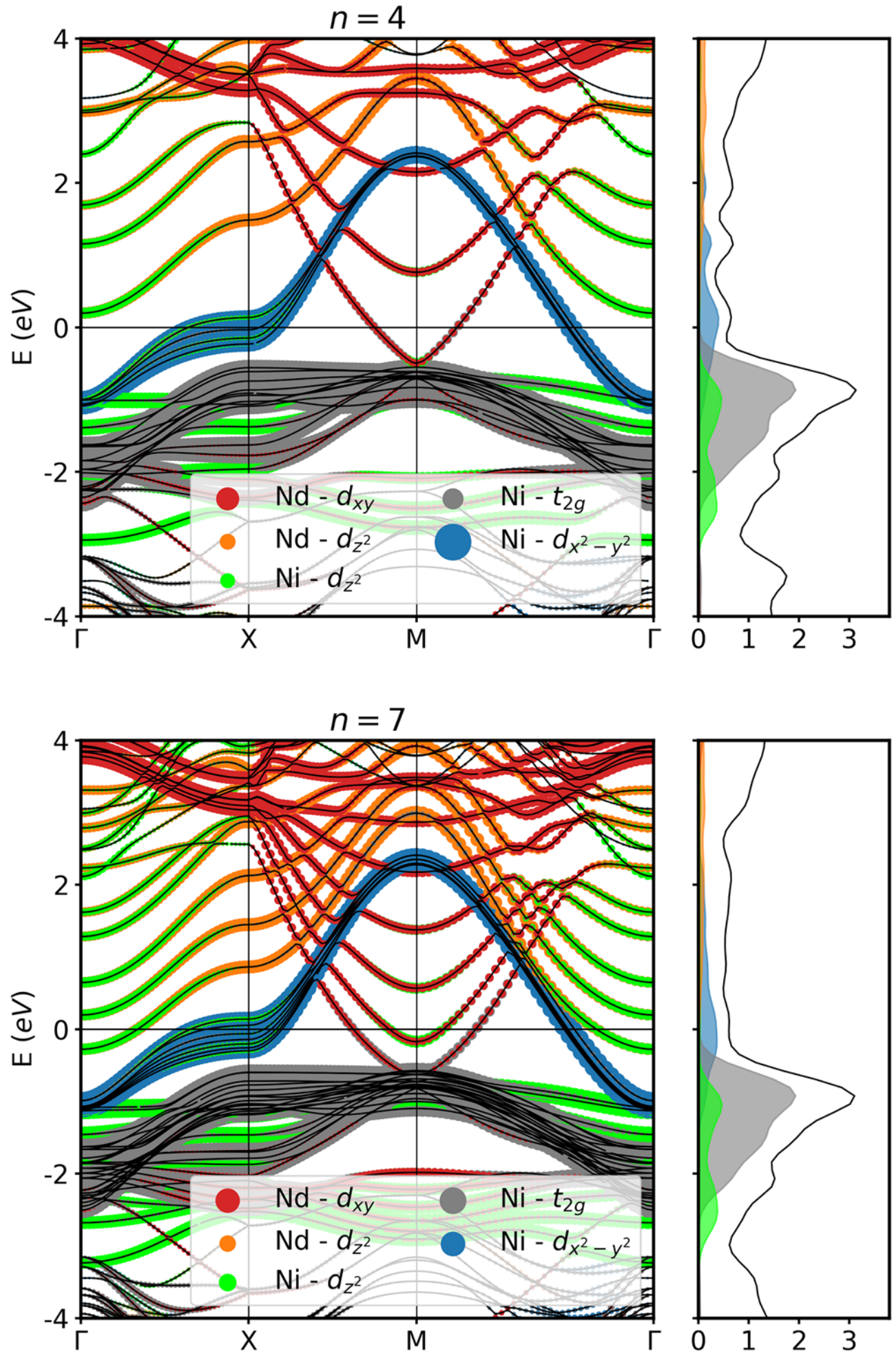

**Fig. 2 | DFT orbital-resolved band structure of $Nd_5Ni_4O_{10}$ (top) and $Nd_8Ni_7O_{16}$ (bottom).** In the right panels the corresponding DFT DOS is shown; see Supplementary Note 3 for $n$ = 2, 3, 5, 6 and ∞ layers. The bands are colored with the projection onto the atomic orbitals. Red, gray, orange, blue, and green indicate Nd $d_{xy}$, Ni $t_{2g}$, Nd $d_{z^2}$, Ni $d_{x^2-y^2}$, and Ni $d_{z^2}$ orbitals, respectively. The DOS is indicated in units of states/eV/Ni; projections onto atomic orbitals follow the same color coding as the bands.

### Dynamical mean-field theory

Due to the partially occupied 3$d$ states, specifically the roughly half-filled Ni 3$d_{x^2-y^2}$ orbital, electronic correlations beyond DFT are important, and we take these into account, on a first level, by DMFT[44,45]. To this end, we perform Wannier projections of the band structure onto a model of Ni 3$d$ and Nd 5$d$ states. Next, we supplement the Wannier-projected Hamiltonian by a local intra-orbital Coulomb interaction of $U$ = 4.4 eV (2.5 eV) and a Hund's exchange $J$ = 0.65 eV (0.25 eV) for the five Ni (Nd) 3$d$ (5$d$) orbitals, as calculated previously in constrained random phase approximation (cRPA)[33,34] and used throughout our work[34,40,41,43]. For details of the Wannier projection and the hopping parameters, see the Supplementary Note 2.

Constructing accurate Wannier projections for large-$n$ superlattices becomes very cumbersome. Yet, such models are also dispensable,

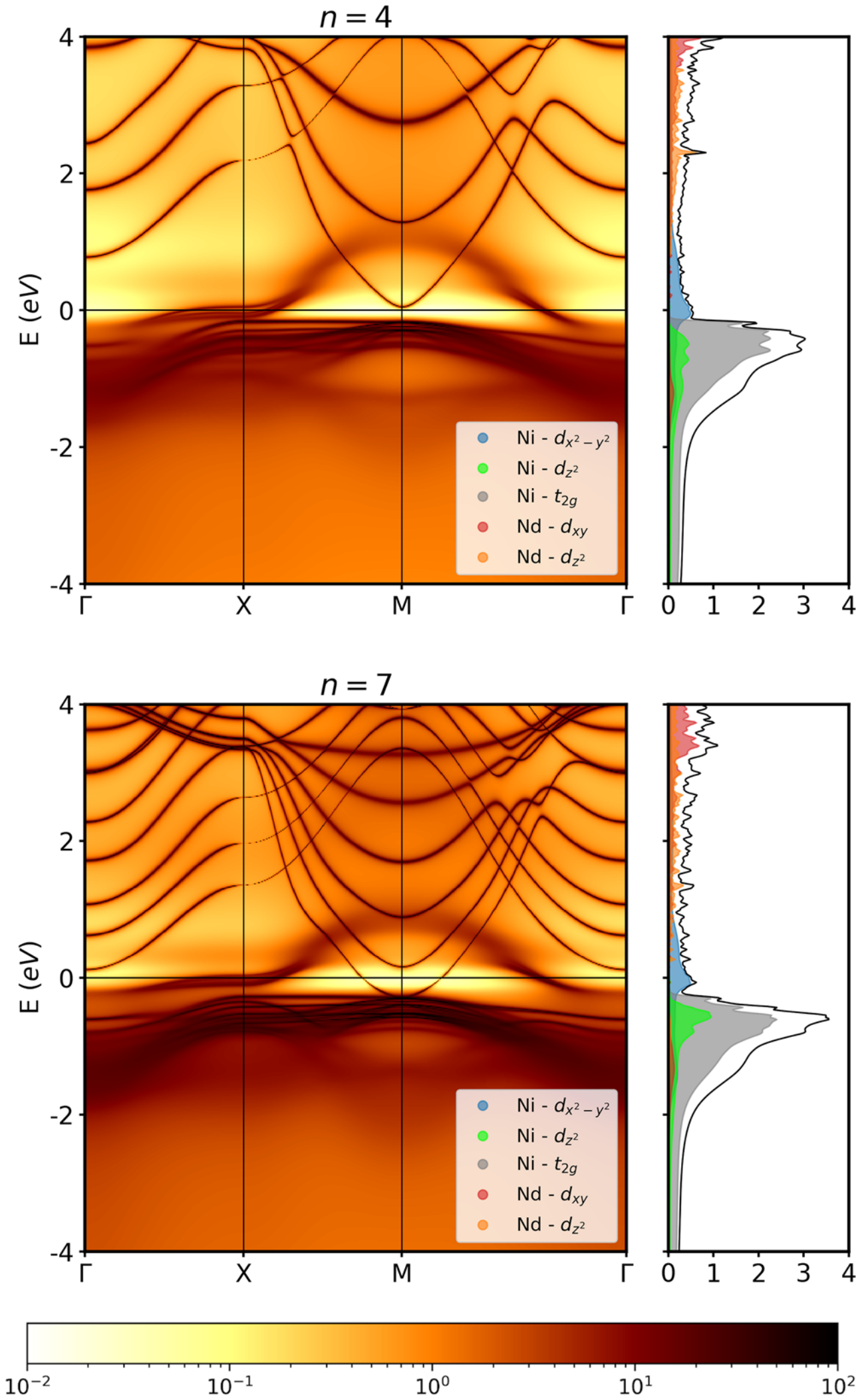

**Fig. 3 | DMFT spectral functions of $Nd_5Ni_4O_{10}$ (top) and $Nd_8Ni_7O_{16}$ (bottom).** In the right panels the corresponding DMFT DOS is shown; calculations are at room temperature; see Supplementary Note 4 for $n = 2, 3, 5, 6, \infty$ layers.

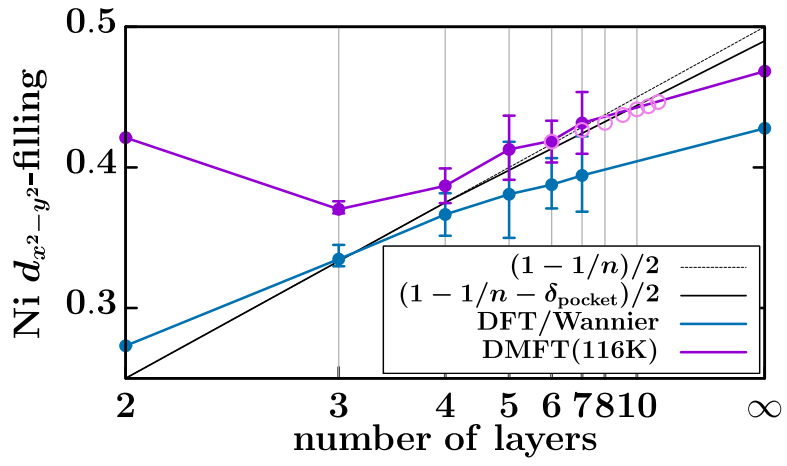

**Fig. 4 | Average Ni $d_{x^2-y^2}$ occupation as a function of the number of layers *n*.** Compared are DMFT at 116 K (which is almost identical to 300 K) to DFT to the formal valence count with and without including the self-doping through the A pocket. Open circles for $n \geq 6$ indicate, as in Fig. 1, an infinite-layer calculation with hole doping $1/n$.

because in this limit the physics of finite-layer nickelates hardly differs from infinite-layer $NdNiO_2$. Hence, for $n \geq 8$, we restrict ourselves to five–plus–five DMFT infinite-layer calculations for one formula unit and a total filling of $(9 - 1/n)$ of all ten orbitals. These "infinite-layer" results are indicated in Fig. 1 (above) and Fig. 4 (below) through open circles and continue those of the actual finite-layer calculations quite smoothly with only a minor kink.

Figure 3 shows the DMFT **k**-resolved and **k**-integrated spectral function, $A(\mathbf{k}, \omega)$ and $A(\omega)$, that can be directly compared, respectively, with the DFT-calculated band structure and DOS. We see an upshift of the $\Gamma$ pocket above the Fermi energy also for $n = 7$ in DMFT. For $n = 4$ now also the A pocket is shifted above the Fermi level, while it still crosses the Fermi energy for $n = 7$. Only the Ni $d_{x^2-y^2}$ orbital that crosses the Fermi level has a strong quasiparticle renormalization $Z = 0.32$ for $n = 4$ and $Z = 0.26$ for $n = 7$, which also causes a finite-life-time broadening compared to the bare DFT dispersion. The renormalizations $Z$ of all other orbitals remain close to one, see Supplementary Note 4. Further, we see some broadened lower Hubbard band around $-0.8$ eV which comprises all Ni $d$ orbitals.

For determining the proper number of electrons in the Ni $d_{x^2-y^2}$ orbital, the filling of the A pocket needs to be taken into account. In the Supplementary Note 5, we determine the DMFT hole doping due to the A pocket to be roughly $\delta_{\rm pocket} \approx 0.02 - 0.08 \times 1/n$ for $n > 4$ and zero for $n \leq 4$ where there is no A pocket.

Figure 4 compares the filling of the Ni $d_{x^2-y^2}$ orbital with this hole doping $(1 - 1/n - \delta_{\rm pocket})/2$ to the DMFT occupation of the $d_{x^2-y^2}$ orbital, where the factor 1/2 accounts for the spin. For larger $n$ both well agree, though the infinite-layer compound somewhat deviates from our $\delta_{\rm pocket}$ fitted to finite-layer nickelates. For $n \lesssim 4$, on the other hand, we see deviations in DMFT. The reason for these deviations is two-fold: (i) There is some, actually quite small admixture of the Ni $d_{x^2-y^2}$ orbital with the other Ni and Nd orbitals. Both give opposite effects for the filling of Ni $d_{x^2-y^2}$ orbital and essentially cancel for $n > 4$. When the number of layers is smaller this is not the case any more because the Nd layers adjacent to the stacking faults do not contribute in this admixture, see Supplementary Note 4 for a detailed discussion. (ii) For $n = 2$, the Ni $d_{z^2}$ orbital crosses the Fermi level in DMFT (but not in DFT; see Supplementary Note 4). Its depopulation allows the Ni $d_{x^2-y^2}$ orbital to stay closer to half-filling.

This also means that for $n = 2$ we have multi-orbital physics involving both Ni $d_{z^2}$ and $d_{x^2-y^2}$ similar as for the Ruddlesden-Popper nickelate $La_3Ni_2O_7$[17]. For $n > 2$, on the other hand, there is neither a Ni $d_{z^2}$ Fermi surface nor a $\Gamma$ pocket in DMFT, only a Nd-derived A pocket, in agreement with ARPES experiments for the infinite-layer limit $n \to \infty$[36,37].

Figure 5 further shows the layer- and orbital-resolved DMFT occupation for $n = 7$, with the aforementioned orbital admixture leading to a Ni $d_{z^2}$ occupation around 0.9 instead of 1, some Nd $d_{z^2}$ occupation and a Nd $d_{xy}$ occupation larger than what to expected from the size of the A pocket alone. Most importantly, however, for the Ni $d_{x^2-y^2}$ orbital we observe a quite strong layer dependence. There are sizably more holes in the layers adjacent to the extra O at the stacking fault (red layer) that is responsible for the hole doping in the first place.

## Dynamical vertex approximation

By construction, DMFT accounts only for local correlations. To include non-local correlations and calculate the superconducting $T_c$, the Hamiltonian with $n \times 5$ Ni $3d$ and $(n + 1) \times 5$ Nd $5d$ orbitals is way beyond the current technical limitations. However, the DMFT clearly shows the prevalence of the Ni $3d_{x^2-y^2}$ orbital; and the Nd-derived band(s) crossing the Fermi level around A (and M), do not hybridize in the bulk with the Ni $3d_{x^2-y^2}$ orbital. Also for the finite-layer nickelates this hybridization is weak. Hence, we describe the superconducting properties of the compound as in infinite-layer nickelates[34], i.e., we restrict ourselves to a single $3d_{x^2-y^2}$ orbital per $NiO_2$ layer and do an additional Wannier projection onto this single orbital. The hopping parameters of this one-Ni-$d_{x^2-y^2}$-orbital model are given in the Supplementary Note 2; and the Hubbard interaction $U$ for this single orbital must be somewhat smaller due to additional screening. We use the same cRPA-estimated $U = 8t$ as in Ref. 34.

Given the rather weak inter-layer hopping $t_z$, we then solve a two-dimensional Hubbard model with D$\Gamma$A[34,46–48]. We find $d$-wave superconductivity induced by antiferromagnetic spin fluctuations and the phase diagram has been already presented in Fig. 1. As different layers exhibit slightly different hoppings and dopings, see Fig. 4, we give – as a fair estimate of the error bar – the minimal and maximal $T_c$ for three dopings: average DMFT $d_{x^2-y^2}$ occupation; layer with optimal DMFT $d_{x^2-y^2}$ occupation; and $(1 - 1/n - \delta_{\mathrm{pocket}})/2$. The in-plane hopping parameters are quite constant for all $n$ and all layers, see Supplementary Note 2, and well described by $t = 420$ meV, $t' = -0.25t$, $t'' = 0.12t$.

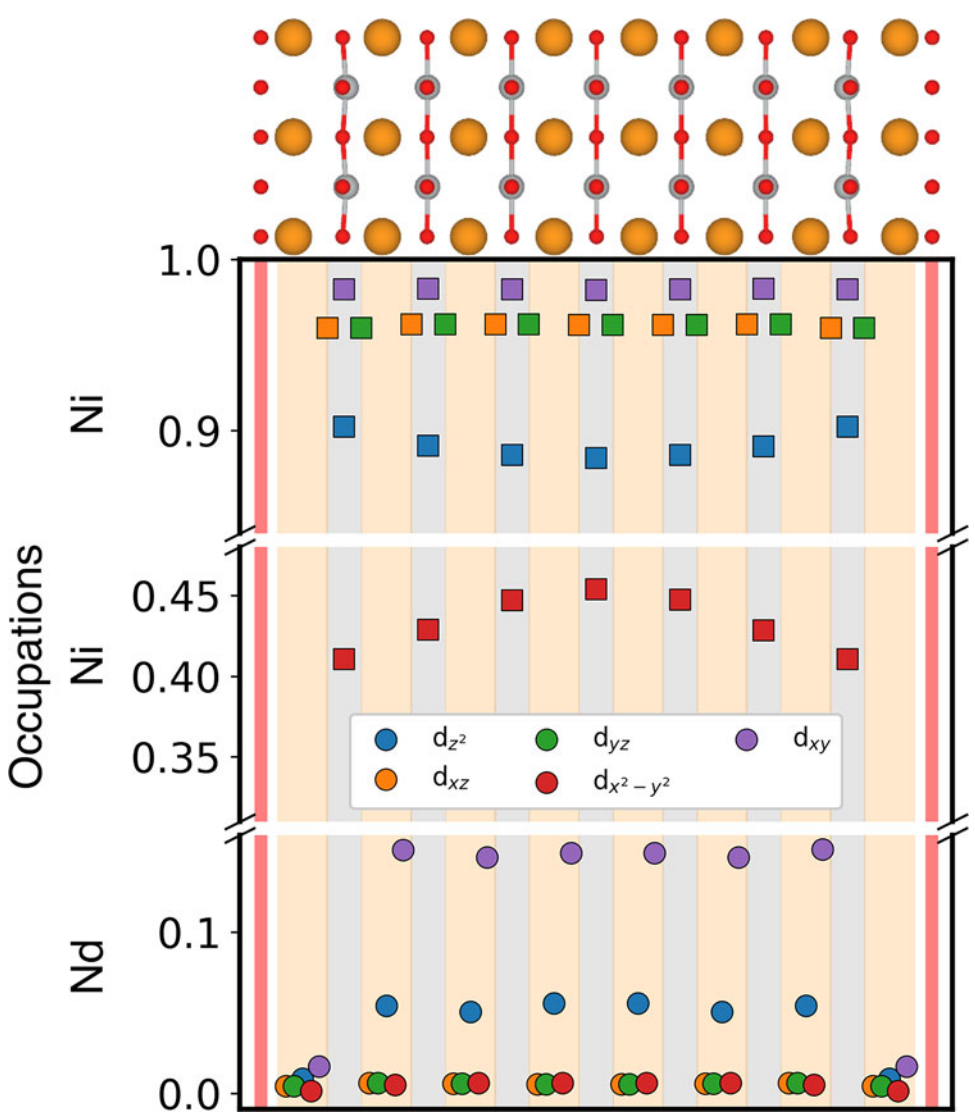

**Fig. 5 | DMFT orbital occupations for each Ni and Nd atom of $Nd_8Ni_7O_{16}$.** See Supplementary Note 4 for other $n$. The top of the figure shows the crystal structure with Nd, Ni and O atoms displayed in orange, gray, and red, respectively. The same color code is used in the bottom part for the background stripes to visualize the Nd, $Ni(O_2)$, and stacking-fault O planes. Orbital occupations of Ni and Nd are displayed in their respective stripe and represented with squares and circles, respectively, with the color code for the different 3$d$ respectively 5$d$ orbitals given in the legend box. Note the broken $y$-axis with the almost empty Nd, half-filled Ni $d_{x^2-y^2}$, and nearly occupied other Ni orbitals.

## Discussion

Using state-of-the-art DFT, DMFT, and D$\Gamma$A we have calculated the superconducting phase diagram of finite-layer nickelates $Nd_{n+1}Ni_nO_{2n+2}$ on a $NdGaO_3$ substrate. In agreement with preliminary experimental results[18], we obtain superconductivity for $n \approx 4$–9 layers, where we have mimicked $n \geq 8$ by a hole-doped infinite-layer Wannier Hamiltonian. Compared to the first infinite-layer experiments,[6,18] the calculated $T_c$ values are somewhat higher. However, the timeline of reported $T_c$ in infinite-layer nickelates makes us hopeful that further experiments and improved synthesis will enhance $T_c$ also in finite-layer nickelates.

Only the Ni $d_{x^2-y^2}$ orbital and the A pocket cross the Fermi energy. A somewhat unexpected result is that the hopping parameters for the Ni $d_{x^2-y^2}$ orbital do *not* depend on the layer index nor on the number of layers $n$. This is a consequence of growing the nickelates on a $NdGaO_3$ substrate, thus fixing the in-plane lattice constant, and the tilting at the stacking fault is, in hindsight, not large enough for changing the hopping parameters substantially. This practical universality of the hopping parameters is very different from the cuprates that are synthesized as bulk materials and are thus much more versatile in their in-plane lattice constant.

However, there is quite a strong layer dependence of the Ni $d_{x^2-y^2}$ orbital occupation. This is relevant for the superconducting $T_c$. Within the constraints of what is presently feasible, we take this layer-dependent doping into account by calculating $T_c$ in D$\Gamma$A for a two-dimensional single-$d_{x^2-y^2}$-orbital Hubbard model, with the error bars in Fig. 1 reflecting in particular the uncertainties and layer-dependencies of the doping of the Ni $d_{x^2-y^2}$ orbital. This uncertainty results in a particular large error bar for $n = 4$. We feel that, nonetheless and with this caveat in mind, our calculation of the phase diagram is accurate and reliable, without any adjusted parameters.

## Methods

In this section, we summarize the computational methods employed. The interested reader can find additional information in[43] (including its Supplementary Information) and[34,49]; data and input files for the whole set of calculations in the associated data repository https://doi.org/10.48436/d608p-bba49.

Density functional theory calculations were performed using the Vienna ab-initio simulation package (VASP)[50,51] using projector-augmented wave pseudopotentials and Perdew-Burke-Ernzerhof exchange correlation functional adapted for solids (PBESol)[52,53], with a cutoff of 1000 eV for the plane wave expansion. Integration over the Brillouin zone was performed over a grid with a uniform spacing of 0.25 $Å^{-1}$ and a Gaussian smearing of 0.20 eV. Wannierization was performed using `wannier90`[54].

The crystal structures structures were relaxed first by fixing the in-plane lattice parameter to the $NdGaO_3$ value (reference substrate, 3.83 Å). Then the internal coordinates were relaxed along with the $c$ axis as in ref. 43.

DMFT calculations were performed using `w2dynamics`[55] using continuous-time quantum Monte Carlo simulations in the hybridization expansion (CTHYB) as an impurity solver[56,57], a Kanamori interaction with the values of $U$ and $J$ as detailed in the main text has been taken at room temperature (300 K) and 116 K (corresponding to 0.01 eV). The fully localized limit was used as double-counting scheme, as introduced in[58]. The self-energies were continued from the Matsubara to the real-frequency axis using the package `ana_cont` and the "chi2kink" method as described in ref. 59.

For obtaining $T_c$, we employed the ladder D$\Gamma$A with the $\lambda$ correction as done before in refs. 34,60–62. To evaluate $T_c$ of the target fillings for drawing the phase diagram, we interpolated (extrapolated) the D$\Gamma$A results for Ni $d_{x^2-y^2}$ fillings $n = 0.775$, 0.80, 0.825, 0.85, 0.875, 0.90, which were obtained from ref. 34 (See Supplementary Note 6 for more details).

## Data availability

The raw data for the figures reported, along with input and output files, is available at the TU Wien repository https://doi.org/10.48436/d608p-bba49.

## Acknowledgements

We thank L. Si and P. Worm for helpful discussions. We further acknowledge funding by the project SuperNickel, jointly funded by the Austrian Science Funds (FWF) Grant DOI 10.55776/I5398 and the Deutsche Forschungsgemeinschaft (DFG) Grant No. 465000489, and Grant-in-Aids for Scientific Research (JSPS KAKENHI) Grants No. JP23H03817, and No. JP24K17014. The calculations have been done in part on the Vienna Scientific Cluster (VSC) and on the GCS Supercomputer SuperMUC at Leibniz Supercomputing Centre (www.lrz.de). A.H. gratefully acknowledges the Gauss Centre for Supercomputing e.V. (www.gauss-centre.eu) for funding this project by providing computing time. For the purpose of open access, the authors have applied a CC BY public copyright license to any Author Accepted Manuscript version arising from this submission.

## Author contributions

S.D.C. performed the DFT calculations and wannierization, A.H. the DMFT calculations, M.K. calculated the DΓA $T_c$, O.J. and K.H. devised and supervised the project, and K.H. did the major part of the writing. All authors discussed and refined the project, contributed to the writing and approved the submitted version.

## Competing interests

The authors declare no competing interests.

## Additional information

**Correspondence** and requests for materials should be addressed to Karsten Held.

# Supplementary Information: Superconducting phase diagram of finite-layer nickelates $Nd_{n+1}Ni_nO_{2n+2}$

Andreas Hausoel ,[1,*] Simone Di Cataldo ,[2,*] Motoharu Kitatani ,[3] Oleg Janson ,[1] and Karsten Held [4]

[1] *Institute for Theoretical Solid State Physics, Leibniz Institute for Solid State and Materials Research Dresden, Helmholtzstr. 20, 01069 Dresden, Germany*
[2] *Dipartimento di Fisica, Sapienza Università di Roma, Piazzale Aldo Moro 5, 00187 Roma, Italy*
[3] *Department of Material Science, University of Hyogo, Ako, Hyogo 678-1297, Japan*
[4] *Institute of Solid State Physics, TU Wien, 1040 Vienna, Austria*

In Supplemental Note 1, we present some details on the structural relaxation for the finite-layer nickleates and quantify the deviation from a planar Ni-O-Ni angle in the $NiO_2$ layers. In Supplemental Note 2, details of the Wannier projection and the tight-binding parameters for the single Ni $3d_{xy}$ orbital are given. In Supplemental Note 3 and Supplemental Note 4 the density functional theory (DFT) and dynamical mean-field theory (DMFT) results of the main text is supplemented by the same figures for the other numbers of layers $n$ considered. In Supplemental Note 5 we calculate the number of electrons in the DMFT A pockets. Finally, in Supplemental Note 6, we briefly recapitulate the formalism for determining $T_c$ using the dynamical vertex approximation (DΓA).

## SUPPLEMENTARY NOTE 1: CRYSTAL STRUCTURES

### A. Details on structural relaxation

The ground-state structures of $Nd_{n+1}Ni_nO_{2n+2}$ for the various number of layers $n = 2 - 7$ and $\infty$ is obtained using the conventional cell. Here, the in-plane lattice parameter is constrained to the one for the NGO substrate, i.e. $a = b = 3.83$Å. Then, the out-of-plane lattice parameter ($c$) is varied in a range of $\pm 5$Å around an initial guessed value of $\frac{\sqrt{3}}{2}(n+1)$, where $n$ is the number of layers. For each value of $c$, the atomic positions is relaxed while the cell is kept fixed. The resulting curve of enthalpy vs $c$ value is then fitted with a fourth-degree polynomial to determine the minimum, which was considered the relaxed value for the $c$ axis. The values of $c$ obtained in this way are then employed to construct the primitive cell. We note that the trend of $c$ axis against $n$ is perfectly linear, with $c = (6.56 \cdot n + 5.80)$Å

| $n$ | $c$ (Å) |
|---|---|
| 2 | 18.9 |
| 3 | 25.5 |
| 4 | 32.1 |
| 5 | 38.7 |
| 6 | 45.2 |
| 7 | 51.7 |

Supplementary Table S1. $c$ axis values as a function of $n$ obtained from the DFT relaxation.

There is considerable Ni-O-Ni out-of-plane distortion, visible with the naked eye in Fig. 1 (top) of the main text. In Fig. S1 we here additionally report the actual angle of this distortion for $n = 2 \ldots 7$. There is a large distortion at the stacking fault (the first and last layer), while the central layers are almost planar.

---

* These authors contributed equally.

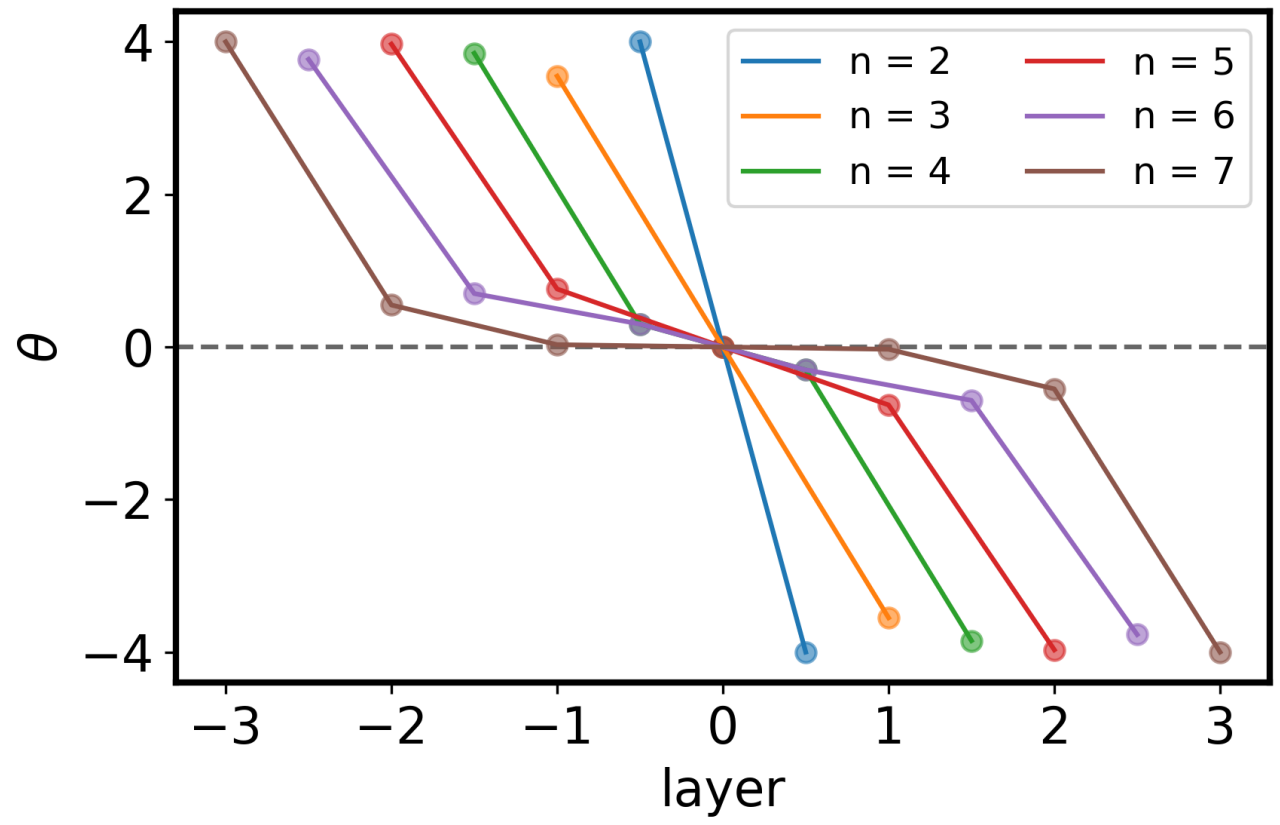

Supplementary Figure S1. Angle formed by the Ni–O bond with respect to the $xy$ plane as a function of the layer number for finite-layer nickelates with $n = 2 - 7$ layers in total. The $x$ axis indicates the layer number, with zero corresponding to the innermost layer.

## SUPPLEMENTARY NOTE 2: WANNIER PROJECTION AND TIGHT-BINDING PARAMETERS

Maximally localized Wannier functions are obtained from the DFT band structure calculated using VASP [1] following the Marzari-Vanderbilt procedure as implemented in Wannier90 [2]. The non-self-consistent calculation of the Kohn-Sham eigenvalues is performed over an $8 \times 8 \times 8$ grid, with an energy cutoff of 1000 eV on the plane waves expansion. The disentanglement of the Wannierization is carried out with the energy windows specified in Tab. S2, with a convergence tolerance of $10^{-10}$Å$^2$ or up to 4200 iterations, with a mix ratio of 0.200. Minimization of the spread is carried out for 1200 iterations. In all cases we find that the spread of the Wannier functions is never larger than 3.0 Å$^2$. In Table S2, we report in detail the outer energy window, the frozen energy window, and the initial projections, for each of the $n$ layers.

| number of layers | `dis_win_min` | `dis_win_max` | `froz_win_min` | `froz_win_max` | `begin_projections` |
|---|---|---|---|---|---|
| Ni d + Nd d; 10×$n$-bands wannierization | | | | | |
| n = 2 | -4.50 | 10.80 | -1.19 | 2.61 | Ni:d, Nd:d |
| n = 3 | -4.87 | 10.13 | -1.66 | 2.14 | Ni:d, Nd:d |
| n = 4 | -4.40 | 9.20 | -1.89 | 1.71 | Ni:d, Nd:d |
| n = 5 | -6.18 | 9.72 | -1.87 | 1.53 | Ni:d, Nd:d |
| n = 6 | -4.73 | 10.57 | -2.22 | 0.58 | Ni:d, Nd:d |
| n = 7 | -3.77 | 10.53 | -1.26 | 1.54 | Ni:d, Nd:d |
| Ni - $d_{x^2-y^2}$; 1×$n$-bands wannierization | | | | | |
| n = 2 | -1.13 | 2.97 | -0.13 | 0.37 | Ni:$d_{x^2-y^2}$ |
| n = 3 | -1.40 | 2.73 | - | - | Ni:$d_{x^2-y^2}$ |
| n = 4 | -1.63 | 2.67 | - | - | Ni:$d_{x^2-y^2}$ |
| n = 5 | -1.37 | 2.92 | - | - | Ni:$d_{x^2-y^2}$ |
| n = 6 | -1.32 | 2.58 | - | - | Ni:$d_{x^2-y^2}$ |
| n = 7 | -2.49 | 4.01 | - | - | Ni:$d_{x^2-y^2}$ |

Supplementary Table S2. Summary of the main Wannierization parameters for the various number of layers $n$. **Note:** to avoid machine or pseudopotential dependency, all the energy values are given with respect to the Fermi energy.

In Tables S3 to S8 we report the hopping parameters for the Wannier projection onto a single Ni $3d_{xy}$ orbital for the $Nd_{n+1}Ni_nO_{2n+2}$ series, with $n$ going from 2 to 7. Here, $t$ is the nearest neighbor hopping, $t'$ the diagonal next-nearest neighbor hopping , $t''$ the next-next-neighbor hopping with two-site distance along the $x$ (or $y$) direction, and $t_z$ nearest neighbor hopping in the $z$-direction between the $NiO_2$ planes.

| n = 2 | $t$ (eV) | $t'$ (eV) | $t''$ (eV) | $t'/t$ | $t''/t$ | $t_z$ (eV) |
|---|---|---|---|---|---|---|
| | -0.41 | 0.10 | -0.06 | -0.24 | 0.14 | -0.06 |
| | -0.41 | 0.10 | -0.06 | -0.24 | 0.14 | |

Supplementary Table S3. Hopping parameters for $n = 2$. The hoppings are listed in the order of the layers (from top to bottom). The interlayer hopping $t_z$ between Ni-d$_{x^2-y^2}$ orbitals in adjacent layers is shown between one layer and the next.

| n = 3 | $t$ (eV) | $t'$ (eV) | $t''$ (eV) | $t'/t$ | $t''/t$ | $t_z$ (eV) |
|---|---|---|---|---|---|---|
| | -0.42 | 0.10 | -0.05 | -0.24 | 0.12 | -0.03 |
| | -0.42 | 0.10 | -0.05 | -0.25 | 0.12 | -0.03 |
| | -0.42 | 0.10 | -0.05 | -0.24 | 0.12 | |

Supplementary Table S4. As in Table S4 but for $n = 3$.

| n = 4 | $t$ (eV) | $t'$ (eV) | $t''$ (eV) | $t'/t$ | $t''/t$ | $t_z$ (eV) |
|---|---|---|---|---|---|---|
| | -0.41 | 0.10 | -0.05 | -0.24 | 0.13 | -0.03 |
| | -0.42 | 0.10 | -0.05 | -0.25 | 0.13 | -0.03 |
| | -0.42 | 0.10 | -0.05 | -0.25 | 0.13 | -0.03 |
| | -0.41 | 0.10 | -0.05 | -0.24 | 0.13 | |

Supplementary Table S5. As in Table S4 but for $n = 4$.

| n = 5 | $t$ (eV) | $t'$ (eV) | $t''$ (eV) | $t'/t$ | $t''/t$ | $t_z$ (eV) |
|---|---|---|---|---|---|---|
| | -0.41 | 0.10 | -0.05 | -0.24 | 0.13 | -0.03 |
| | -0.41 | 0.10 | -0.05 | -0.25 | 0.12 | -0.03 |
| | -0.41 | 0.11 | -0.05 | -0.25 | 0.12 | -0.03 |
| | -0.41 | 0.10 | -0.05 | -0.25 | 0.12 | -0.03 |
| | -0.41 | 0.10 | -0.05 | -0.24 | 0.13 | |

Supplementary Table S6. As in Table S4 but for $n = 5$.

| n = 6 | $t$ (eV) | $t'$ (eV) | $t''$ (eV) | $t'/t$ | $t''/t$ | $t_z$ (eV) |
|---|---|---|---|---|---|---|
| | -0.41 | 0.10 | -0.05 | -0.25 | 0.13 | -0.03 |
| | -0.42 | 0.10 | -0.05 | -0.24 | 0.13 | -0.03 |
| | -0.41 | 0.10 | -0.05 | -0.25 | 0.13 | -0.03 |
| | -0.41 | 0.10 | -0.05 | -0.25 | 0.13 | -0.03 |
| | -0.42 | 0.10 | -0.05 | -0.24 | 0.13 | -0.03 |
| | -0.41 | 0.10 | -0.05 | -0.25 | 0.13 | |

Supplementary Table S7. As in Table S4 but for $n = 6$.

| n = 7 | $t$ (eV) | $t'$ (eV) | $t''$ (eV) | $t'/t$ | $t''/t$ | $t_z$ (eV) |
|---|---|---|---|---|---|---|
| | -0.41 | 0.10 | -0.05 | -0.24 | 0.12 | -0.02 |
| | -0.41 | 0.10 | -0.05 | -0.24 | 0.12 | -0.03 |
| | -0.41 | 0.10 | -0.05 | -0.25 | 0.12 | -0.03 |
| | -0.41 | 0.10 | -0.05 | -0.25 | 0.12 | -0.03 |
| | -0.41 | 0.10 | -0.05 | -0.25 | 0.12 | -0.03 |
| | -0.41 | 0.10 | -0.05 | -0.24 | 0.12 | -0.02 |
| | -0.41 | 0.10 | -0.05 | -0.24 | 0.12 | |

Supplementary Table S8. As in Table S4 but for $n = 7$.

## SUPPLEMENTARY NOTE 3: ADDITIONAL DFT RESULTS

In this Supplemental Note, we present in Fig. S2 the DFT band-structure and density of states (DOS) also for $n = 2$, $n = 3$, $n = 5$ and $n = 6$. The DFT results for $n = 4$ and $n = 7$, already presented in Fig. 2 of the main text, are repeated for convenience.

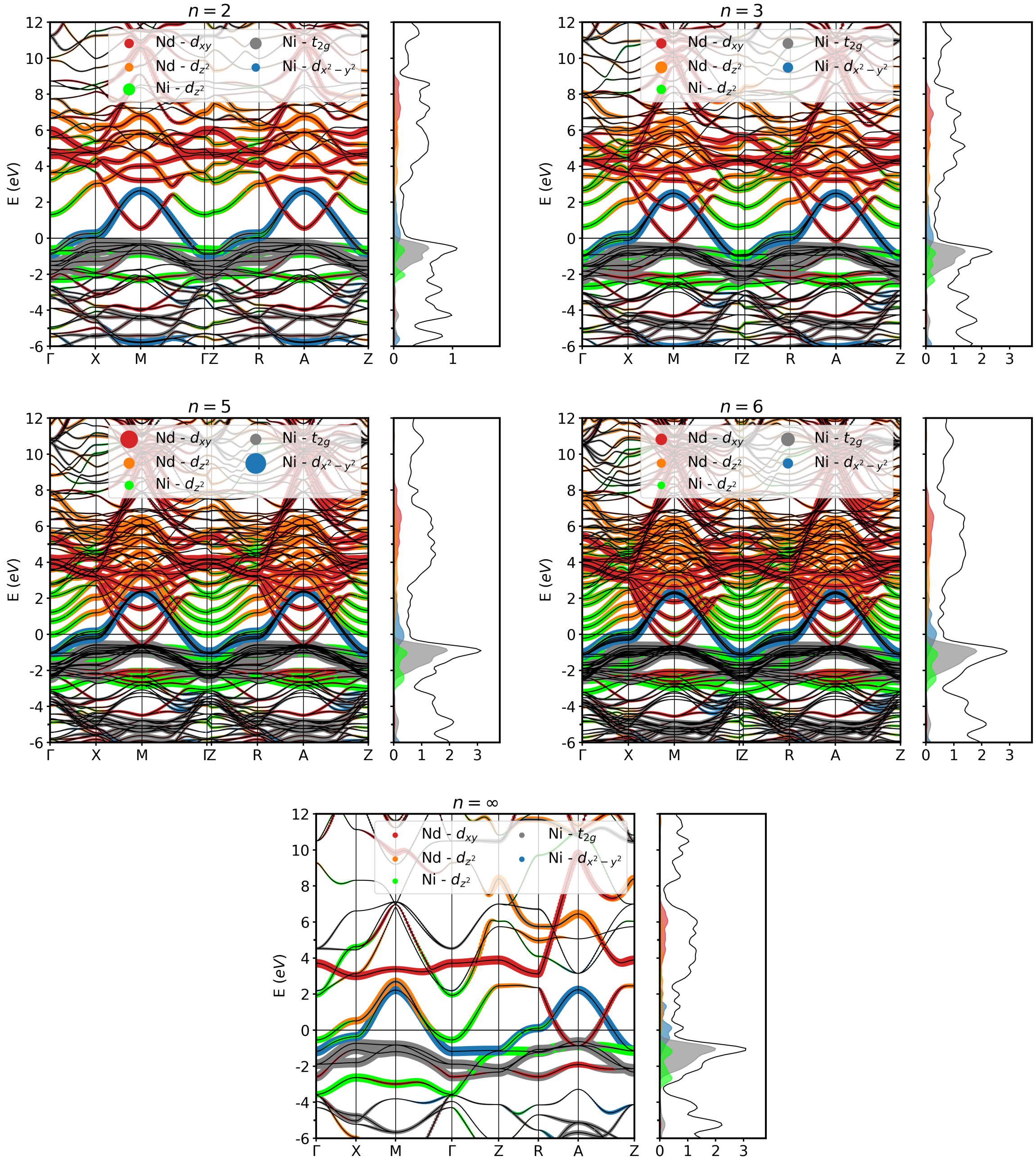

Supplementary Figure S2. DFT band structure for finite-layer nickelates $Nd_{n+1}Ni_nO_{2n+2}$ ($n = 2, 3, 5, 6, \infty$). The Fermi energy is taken as the energy zero. Projections onto different atomic orbitals are shown as thick colored lines, where thickness indicates the strength of the projection.

Further, in Fig. S3 we show the DFT Fermi surface for the finite-layer nickelates considered with $n = 2 \ldots 7$ layers, as well as for the infinite-layer nickelates. With the stacking fault well separating each $n$ layers from the next $n$, there is essentially no $k_z$ dispersion any more. Instead, we get $n$ copies of the Ni 3$d_{xy}$ and $n$+1 copies of the Nd d$_{xy}$ orbital which forms the pocket around the A pocket momentum (see the $k_z = \pi$ plane) for the infinite-layer nickelate. For the finite-layer nickelates, this pocket just emerges for $n = 2$ as a tube instead of a pocket, comprising the A and M point. For $n = 7$ a second tube emerges.

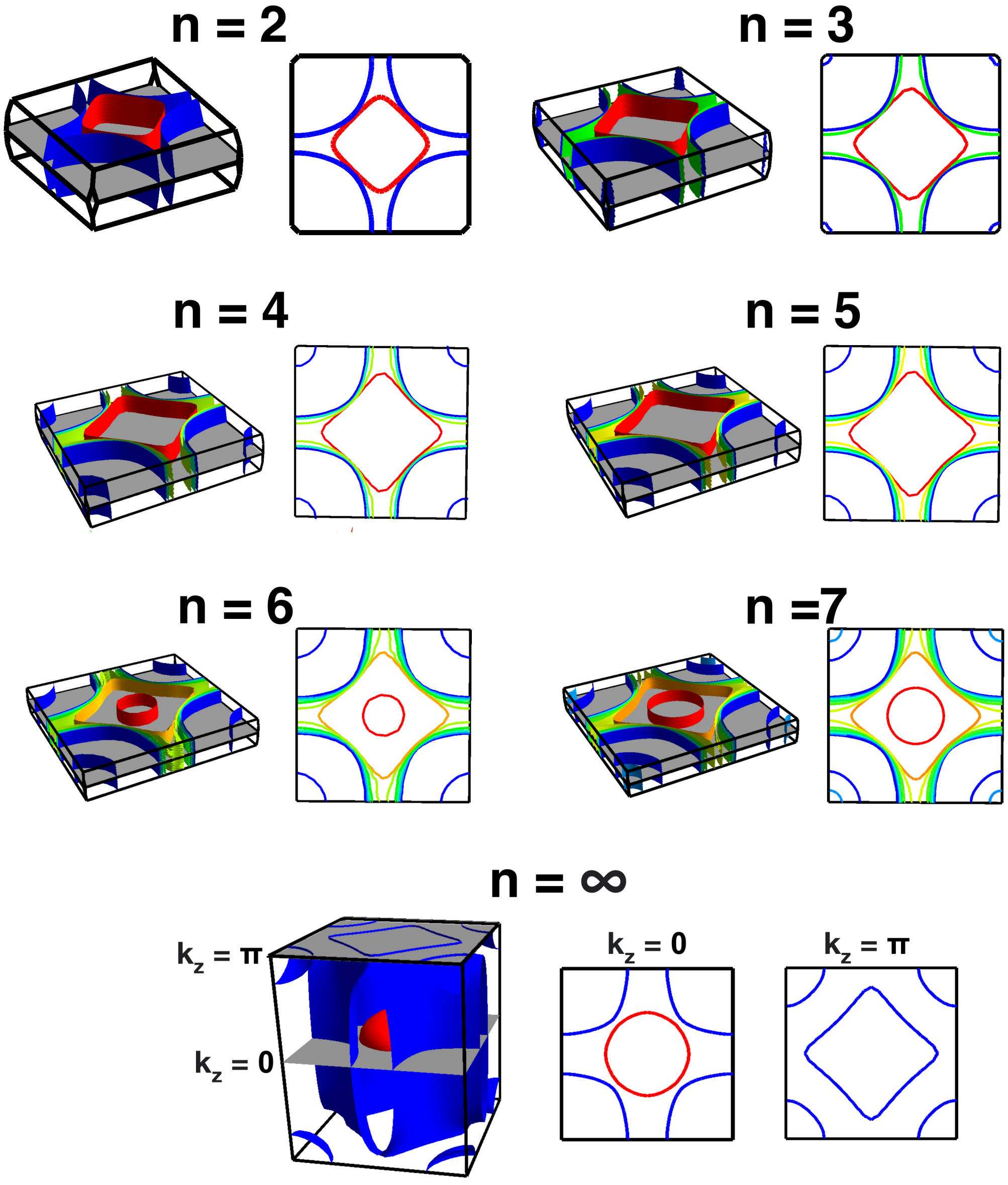

Supplementary Figure S3. DFT-calculated Fermi surfaces for different number of layers $n$.

As for the Ni $3d_{xy}$ Fermi surface, we have for the infinite-layer case in Fig. S3, a hole-like Fermi surface for $k_z = 0$ and an electron-like for $k_z = \pi$. Concomitantly, we have both hole-like and electron-like Fermi surfaces for finite $n$, with the second electron-like Fermi surface appearing for $n = 5$.

Further, in Fig. S4 we show the layer-resolved bandstructure and partial DOS. The overall picture is that the stacking fault perturbs particularly the $z$ direction for the outer layers. This is quite intuitive if one considers that it is only in that direction that the stacking fault breaks symmetry and the outer layers are most affected. The effect on Ni $3d_{x^2-y^2}$ is that the inner layers have more weights on the outer two bands in the four-band strip between $\Gamma$ and $X$. In the DOS this results in a split of the van Hove peak into two peaks just above and below the Fermi energy.

In Fig. S5 we also show the DFT band structure for a heterostructure comprising $n = 3$ and $n = 4$ slabs, with orbital projections resolved with respect to both atomic orbitals and slab. One can see that the bands of different slabs are "independent", i.e. there are no mixed partial projections for the two slabs.

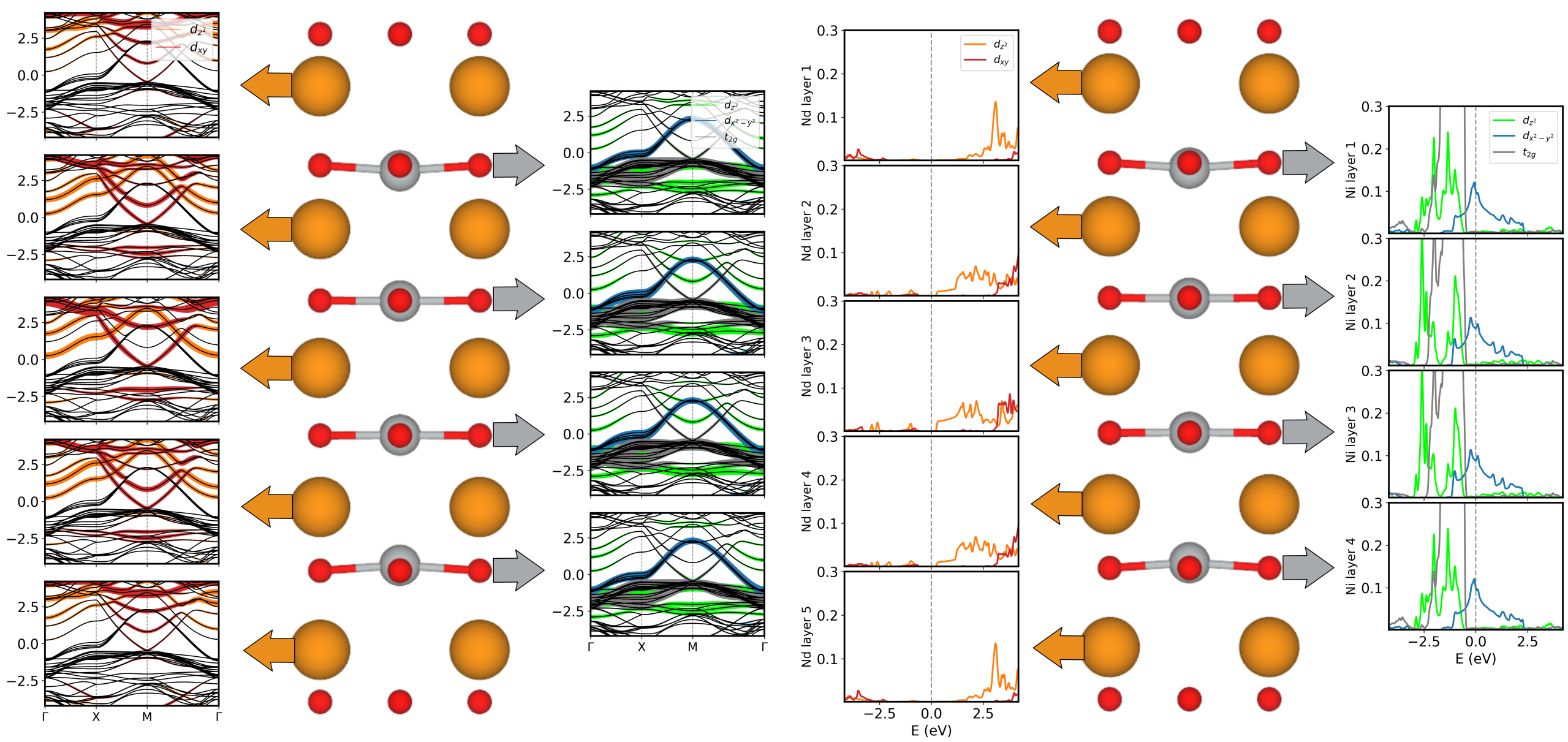

Supplementary Figure S4. Layer-resolved band structures (left) and partial DOS (right) for $n = 4$.

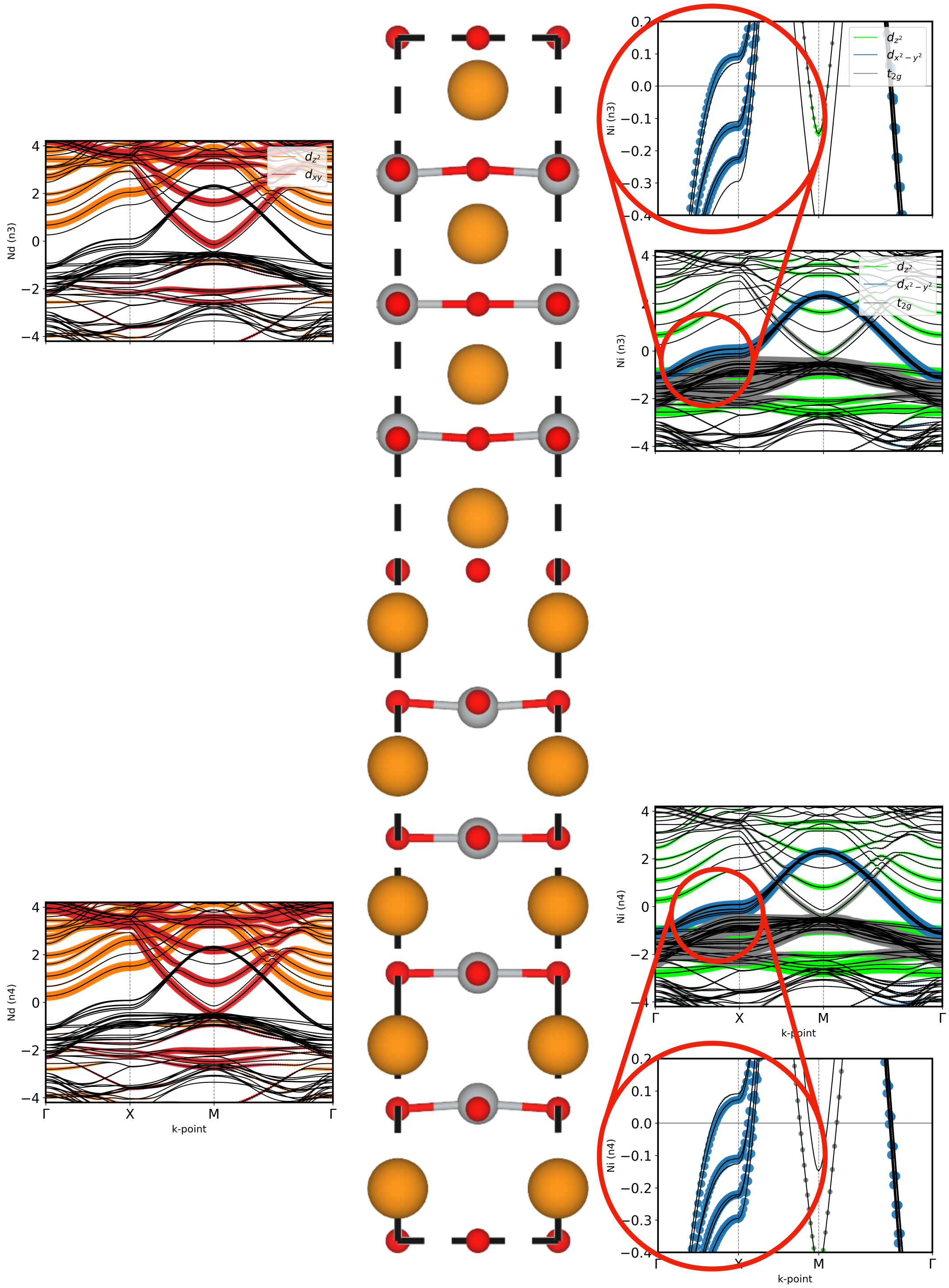

Supplementary Figure S5. DFT band structure of a "$n = 3$" + "$n = 4$" heterostructure with slab- and orbital-resolved projection (fat blue bands). Top: projection on the first $n = 3$ slab; bottom: $n = 4$ slab for Nd orbitals (left) and Ni orbitals (right; including a zoom-in).

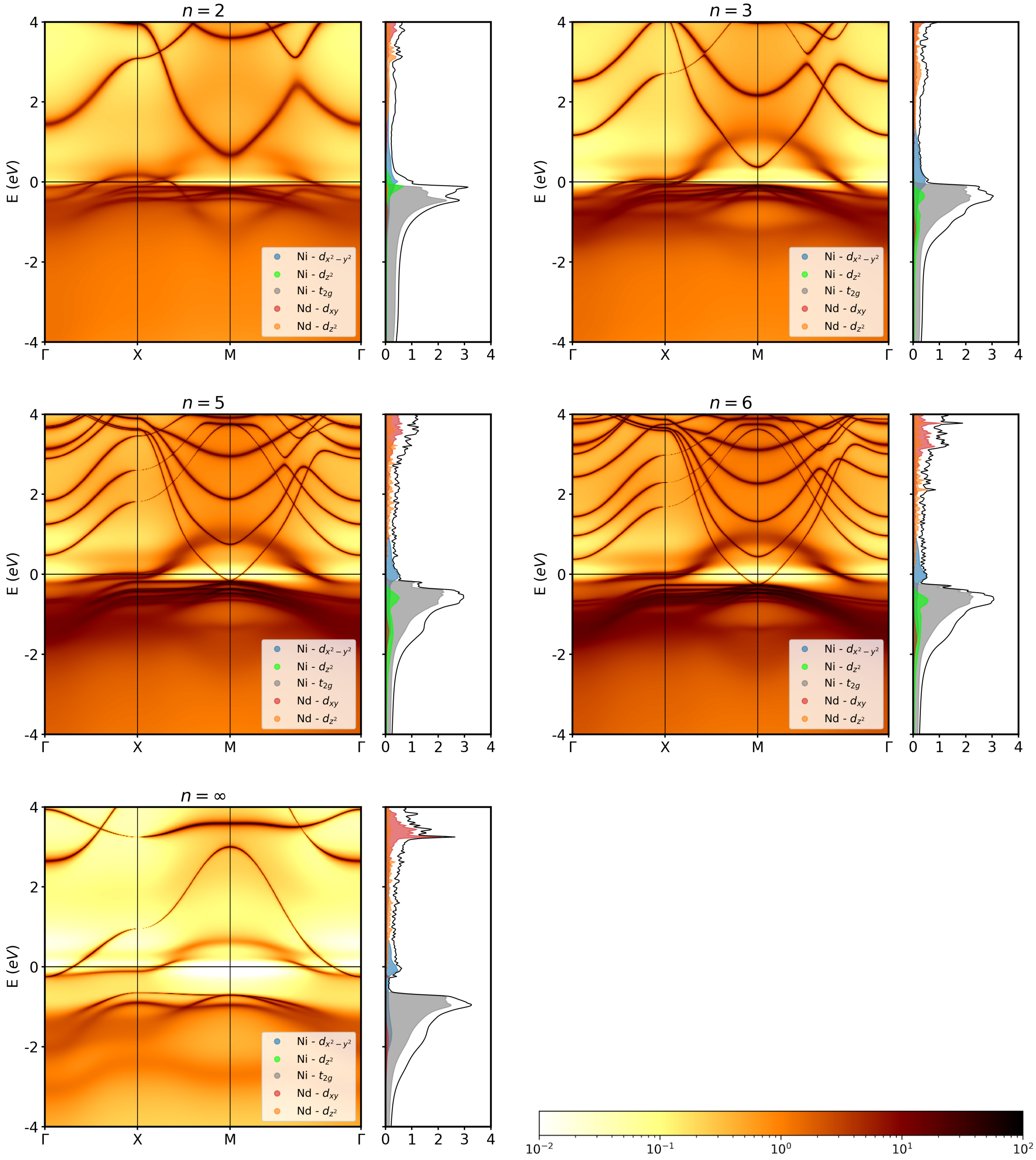

Supplementary Figure S6. DMFT electronic structure and DOS of $Nd_{n+1}Ni_nO_{2n+2}$ for $n = 2, 3, 5, 6, \infty$ at room temperature.

## SUPPLEMENTARY NOTE 4: ADDITIONAL DMFT RESULTS

In Fig. S6 we show, additionally to Fig. 3 of the main text, also the DMFT spectral function and DOS for $n = 2, 3, 5, 6$. For $n = 2$, the hole doping is so large that the DMFT physics turns: the Ni $d_{z^2}$ orbital also crosses the Fermi level and we have actual multi-Ni-orbital physics. For $n > 2$ the qualitative behavior is similar, with as a matter of course more vineyard-like bands for larger $n$. Further we provide in Fig. S7 all calculated quasiparticle renormalizations.

In Fig. S8 we show the occupation for all atoms (layers) and all orbitals also for other $n$ besides the $n = 7$ in Fig. 5 of the main text. Besides the DMFT results, also those of DFT, more precisely the DFT-derived Wannier Hamiltonian are shown. We notice for all $n$ a somewhat larger DMFT occupation for the $3d_{xy}$ orbital than in DFT, but the tendency to have more holes in the layers interfacing the stacking fault is the same in DFT and DMFT. For $n = 2$ the difference between DMFT and DFT occupation is strongly enhanced, as the DMFT Ni $d_{z^2}$ orbital crosses the Fermi level.

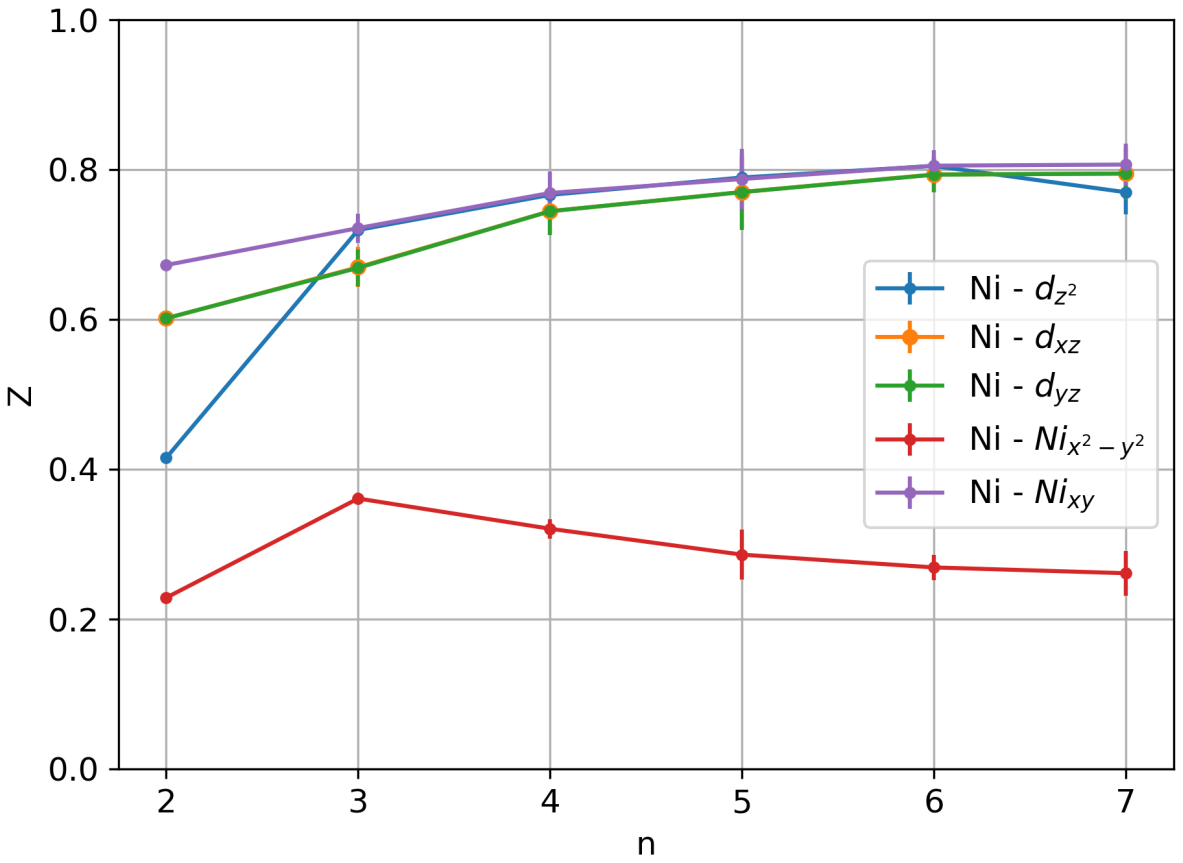

Supplementary Figure S7. Quasiparticle renormalizations of the Ni orbitals in DMFT as a function of the number of layers $n$ at room temperature. Here, orbital 1 to 5 are the Ni $z^2$, $xz$, $yz$, $x^2-y^2$ and $xy$ orbital, respectively. The error bar reflects the variation of $Z$ from layer to layer.

While for $n = 3 \ldots \infty$ no orbital other than the predominantly Ni $3d_{xy}$ quasiparticle band and the Nd-$d_{xy}$-derived pocket cross the Fermi level, this orbital has some (albeit small) admixture with the Ni and Nd other orbitals. This admixture explains the deviations from the occupation $(1 - 1/n - \delta_{\text{pocket}})/2$ of the Ni $3d_{xy}$ band, with $\delta_{\text{pocket}}$ determined in Supplemental Note 5.

For example, in Fig. 2 of the main text or in Fig. S2 one sees that there is also a small Ni $d_{z^2}$ contribution (green) to the predominantly Ni $3d_{xy}$ orbital at and below the X and R point. Vice versa there is a small Ni $3d_{xy}$ admixing to the other Ni bands below the Fermi energy (blue). Furhter, we observe a Ni $t_{2g}$ admixture (grey) to the Nd $d_{xy}$ band around A, and a somewhat larger admixture between Ni $d_{z^2}$ and Nd $d_{z^2}$ (green and orange).

For the orbital- and layer-resolved occupations in Fig. S8, this orbital admixture altogether leads to a Ni $d_{z^2}$ occupation of 0.9 instead of 1, Ni $d_{xz/yz}$ of 0.97; Nd $d_{z^2}$ of about 0.05 and Nd $d_{xy}$ of about 0.15 (which is beyond what to expect from the A pocket size). For the nearly half-occupied Ni $3d_{xy}$ orbital these effects essentially cancel for $n > 4$ in DMFT. In contrast, the DFT occupations in Fig. 4 of the main text lie below $(1 - 1/n - \delta_{\text{pocket}})/2$. In this respect, keep in mind though that the DFT pocket is larger than the plotted $\delta_{\text{pocket}}$ of DMFT and is already present for $n = 3$. The cancellation of the admixture effects in DMFT, is no longer valid for $n \lesssim 3$. The first reason for this is that there is essentially no hybridization with the outermost Nd orbitals: in Fig. S8 all Nd orbitals of the first and last layer are unoccupied, only those in the middle have a finite occupation because of the hybridization. For $n = 2$ a second major factor for the deviation from this analytical expectation is that the DMFT Ni $d_{z^2}$ orbital now also crosses the Fermi level. It is thus much more depopulated, the Ni $d_{x^2-y^2}$ orbital vice versa closer to half-filling.

Further, in Fig. S9 and S10 we show the DMFT self-energies of the Ni and Nd orbitals, respectively, for the 7-layer system as measured directly by continuous-time quantum Monte Carlo simulations in the hybridization expansion at Matsubara frequencies $\omega_n = (2n + 1)\pi T$. The slope of Im$\Sigma$ for $\omega_n \to 0$ directly relates to the quasiparticle renormalization which is largest for the strongly correlated Ni $3d_{x^2-y^2}$ orbital. It is already much smaller for the other Ni orbitals; and for the Nd $5d$ orbitals Im$\Sigma$ is by more than one order of magnitude smaller, reflecting the weak quasiparticle renormalization of the Nd bands.

For R$e\Sigma$, the double counting (DC) correction based on the fully localized limit [3] has been subtracted, as it roughly cancels the Hartreee term, as well as the chemical potential. The real part of the self-energy describes the shift of the bands relative to each other, with electronic correlations pushing the Nd bands somewhat up relative to the Ni ones.

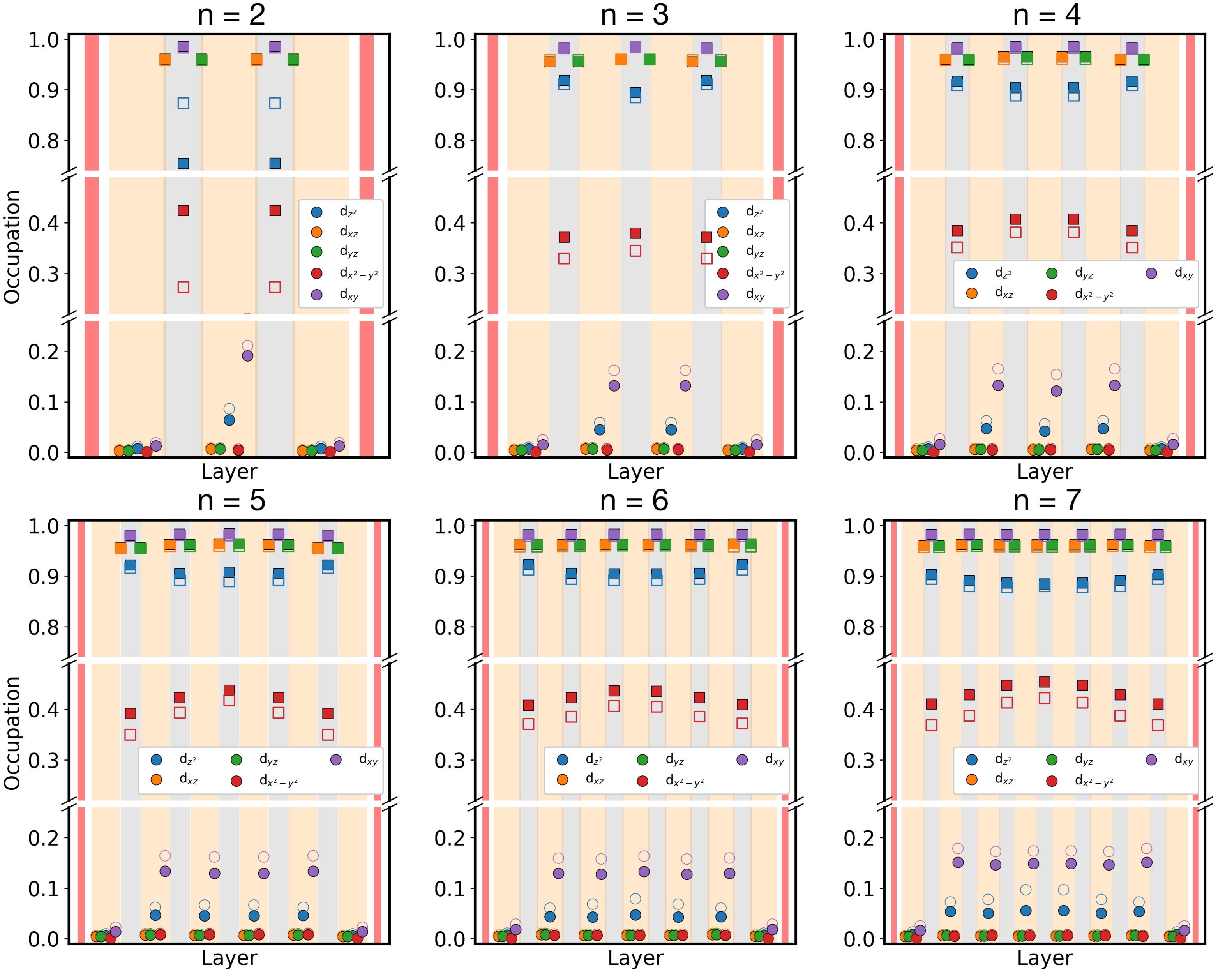

Supplementary Figure S8. Orbital- and layer-resolved Ni and Nd occupations for all the finite-layer compounds studied. The colored orange/grey/red stripes in the background represent the Nd, Ni, and defect O planes. Orbitals relative to Ni and Nd are represented respectively with squares and circles, while filled/hollow symbols indicate the DMFT and DFT (non-interacting Wannier Hamiltonian) result.

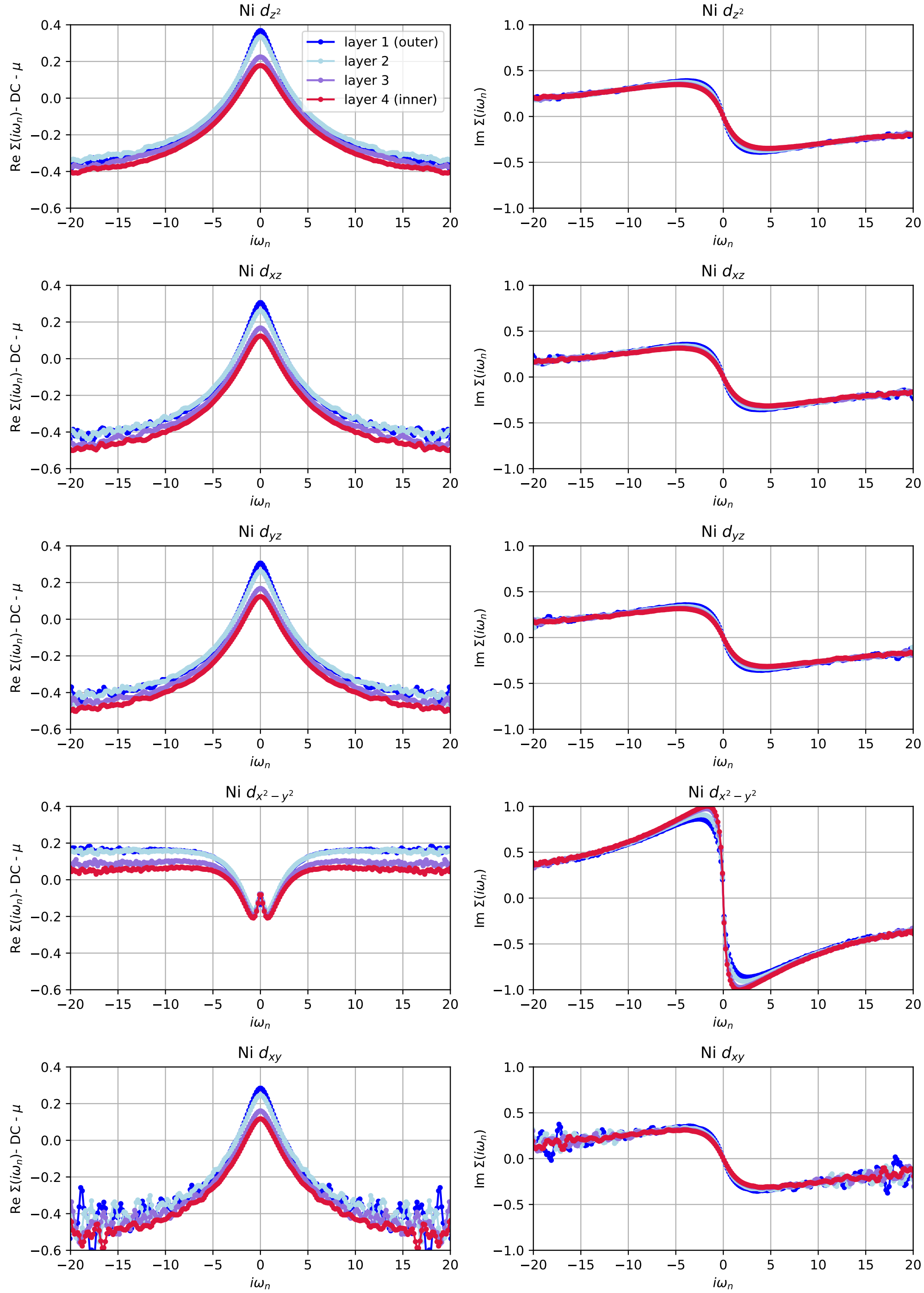

Supplementary Figure S9. Real and imaginary parts of the DMFT self-energies of the Ni 3$d$ orbitals for the 7-layer system.

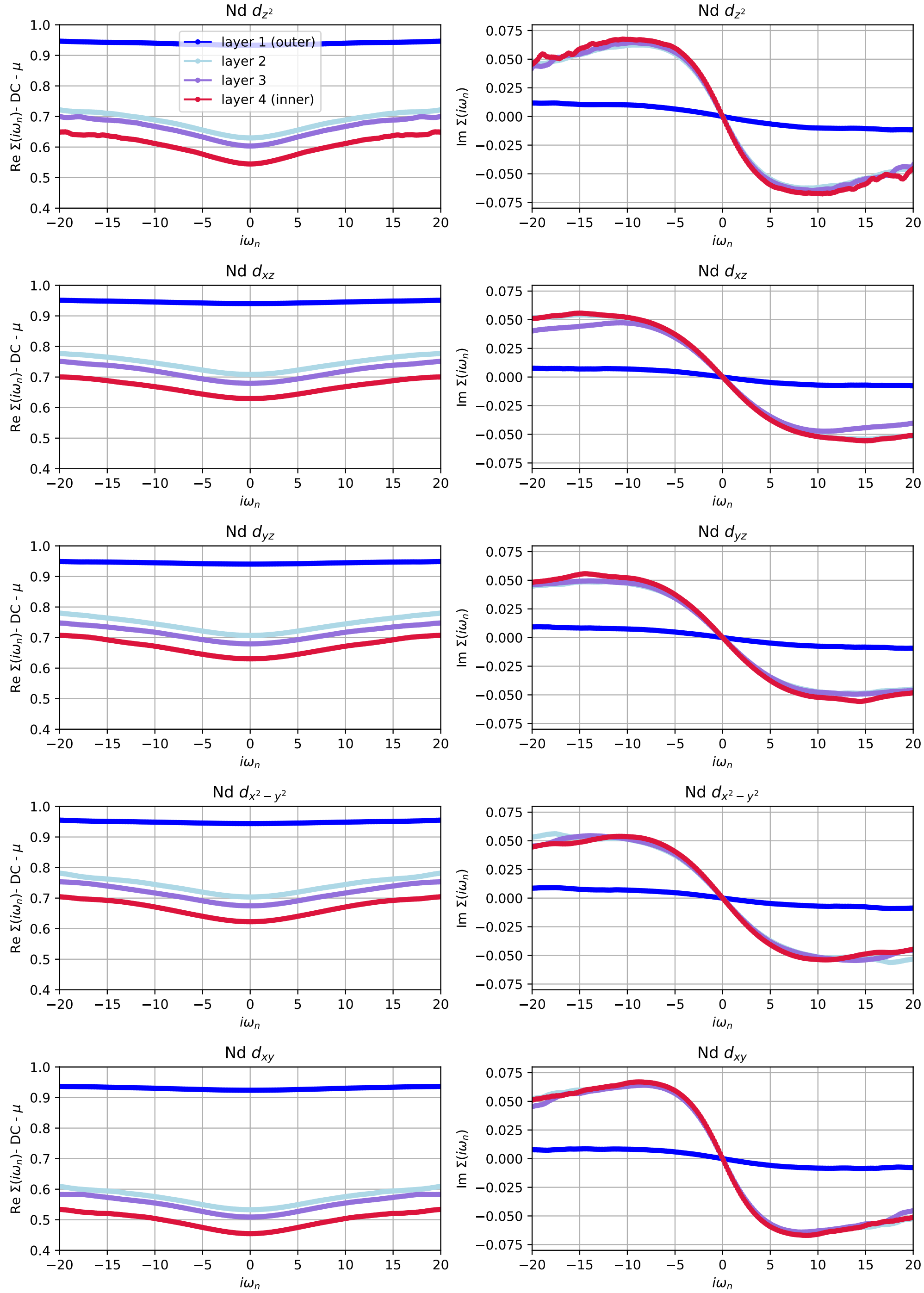

Supplementary Figure S10. Real and imaginary parts of the DMFT self-energies of the Nd 5$d$ orbitals for the 7-layer system.

## SUPPLEMENTARY NOTE 5: ELECTRONS IN THE $A$ POCKET

In Fig. S11 we show the calculated number of electrons in the A pocket as a function of the number of layers $n$. The extrapolated fit (black dashed line) is (rouned to 2 digits) $\delta_{\text{pocket}} = 0.02 - 0.08 \times 1/n$. The calculation of $\delta_{\text{pocket}}$ in DMFT is actually quite involved and outlined in the following. In short, we estimate the number of electrons by the volume of a cylinder representing the Fermi surface of the A–M pocket that also in DMFT is akin to the DFT Fermi surface in Fig. S3 (just shifted).

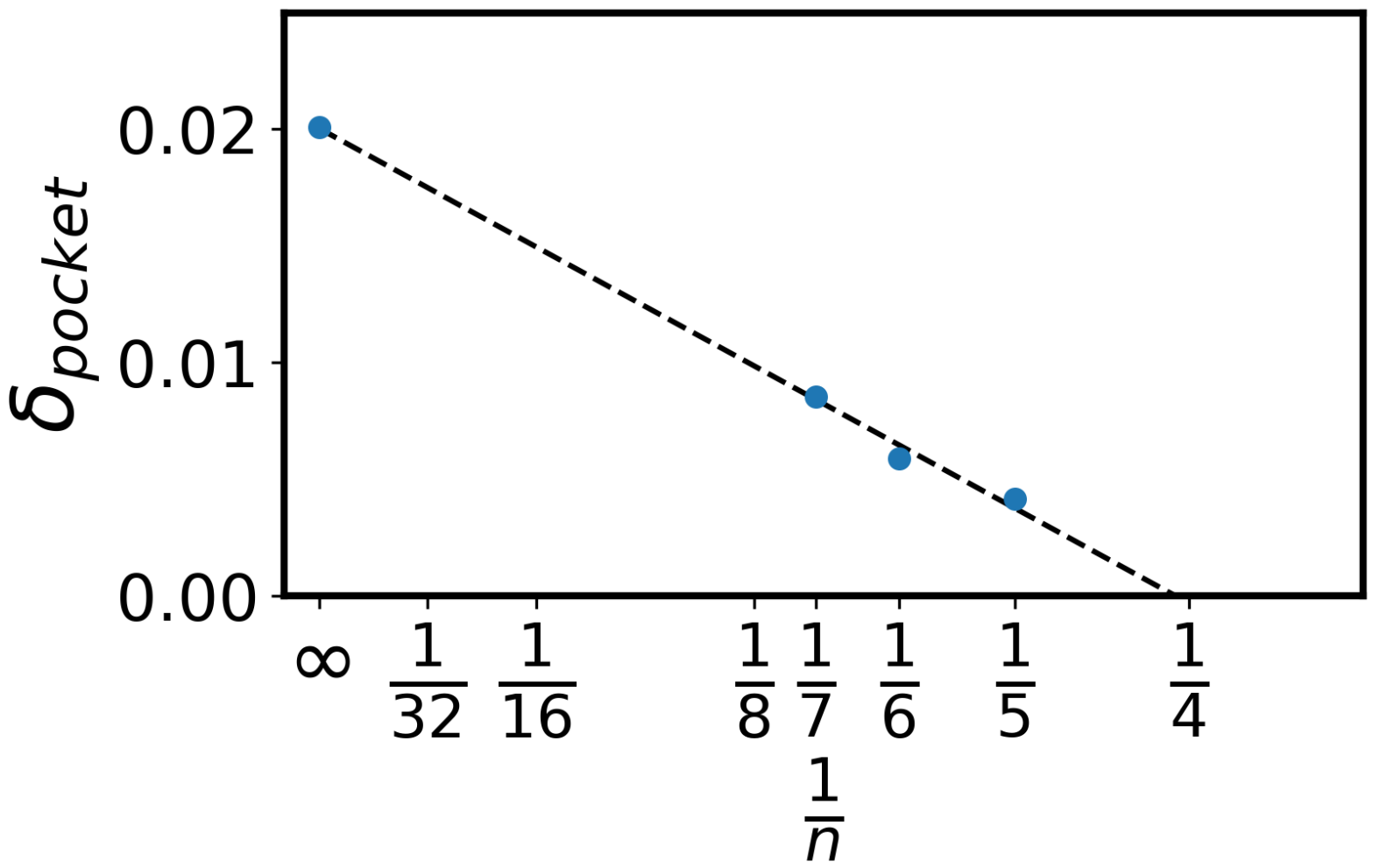

Supplementary Figure S11. Number of electrons in the $A$ pocket as a function of the number of layers, normalized by the number of Ni atoms in the primitive cells. The black line indicates an extrapolated fit.

### A. Analytic calculation

To estimate the occupation of the pocket in the $A$ point, we proceeded as follows. First, we observe (as shown in Fig. S3) that at the DFT level the Fermi surface forming $A$ pocket is a cylinder with its axis going from $A$ to $M$, and this is particularly true for $n \geq 5$. For the infinite-layer case, the pocket is instead rather spherical.

Considering that the number of states per unit volume in the reciprocal space is constant, and that each band contains up to two electrons (counting spin), the number of occupied states in a pocket can be calculated as

$$e_p^- = 2\frac{V_p}{V_{BZ}} \tag{S1}$$

Where $V_p$ is the volume of the pocket, and $V_{BZ}$ is the volume of the Brillouin zone.

*a.* ***Ellipsoidal Fermi surface*** For a Fermi surface in the shape of an ellipsoid (quasi-spherical), the volume enclosed is

$$V_{p,s} = \frac{4}{3}\pi k_{Fx} k_{Fy} k_{Fz} \ , \tag{S2}$$

where $k_{Fx}$, $k_{Fy}$, $k_{Fz}$ are the semi-axes of the spheroid, i.e. a sort of Fermi momentum, generalized to an ellipsoid. Substituting into Eq. (S1) we obtain that the number of states is

$$e_{p,s}^- = \frac{k_F^3 L_x L_y L_z}{3\pi^2} \ , \tag{S3}$$

where $L_x L_y L_z = V$ is the volume of the unit cell in real space.

*b.* ***Cylindrical Fermi surface*** For a cylindrical Fermi surface in a quasi-2D system, the volume enclosed is

$$V_{p,c} = \pi k_F^2 \frac{2\pi}{h} \tag{S4}$$

Where $\frac{2\pi}{h}$ is the height of the Brillouin zone in the shortest direction. Here $k_F$ is to be intended as the radius of the cylinder, i.e. the Fermi momentum for the 2D case. Substituting into Eq. (S1) we obtain that the number of electrons in the A–M spheroid is

$$e^-_{p,c} = \frac{k_F^2 L_x L_y}{2\pi} , \tag{S5}$$

where $L_x L_y = A$ is the area of the top/bottom face of the unit cell in real space.

The $\mathbf{k_F}$ momentum can be identified as the $\mathbf{k}$ value for which the electron dispersion around the $A$ pocket crosses the Fermi energy. This process must take into account the fact that electronic correlations tend to shift the pockets upwards in DMFT. To identify the momentum with the highest possible precision, we adopt the following process:

1. Superimpose the DMFT spectral function $A(\mathbf{k}, \omega)$ with the DFT bands, to find an energy shift $\epsilon^*$ such that the DFT band of the pocket corresponds to the maximum of the spectral function $A(\mathbf{k_F}, \omega = 0)$ (along the $R$–$A$ line). An example for the case of $n = 5$ is shown in Fig. S12.[4]

2. Calculate the DFT bands on an extremely dense array of points between $R$ and $A$

3. Shift the DFT bands by $\epsilon^*$, and find the momentum $\mathbf{k_F}$ for which they cross the Fermi energy.

4. As $A$ is the center of the pocket, the vector $\Delta\mathbf{k} = \mathbf{k_F} - \mathbf{A}$ has modulus equal to the radius of the cylinder (or the ellipsoid semiaxis in the $R$–$A$ direction)

5. **Spheroid only:** Repeat from point 2 along the $A$–$M$ direction to obtain the size of the different semiaxis in that direction.

**Note:** due to the crystal symmetry, the radius of the cylinder must be the same in the $x$ and $y$ direction.

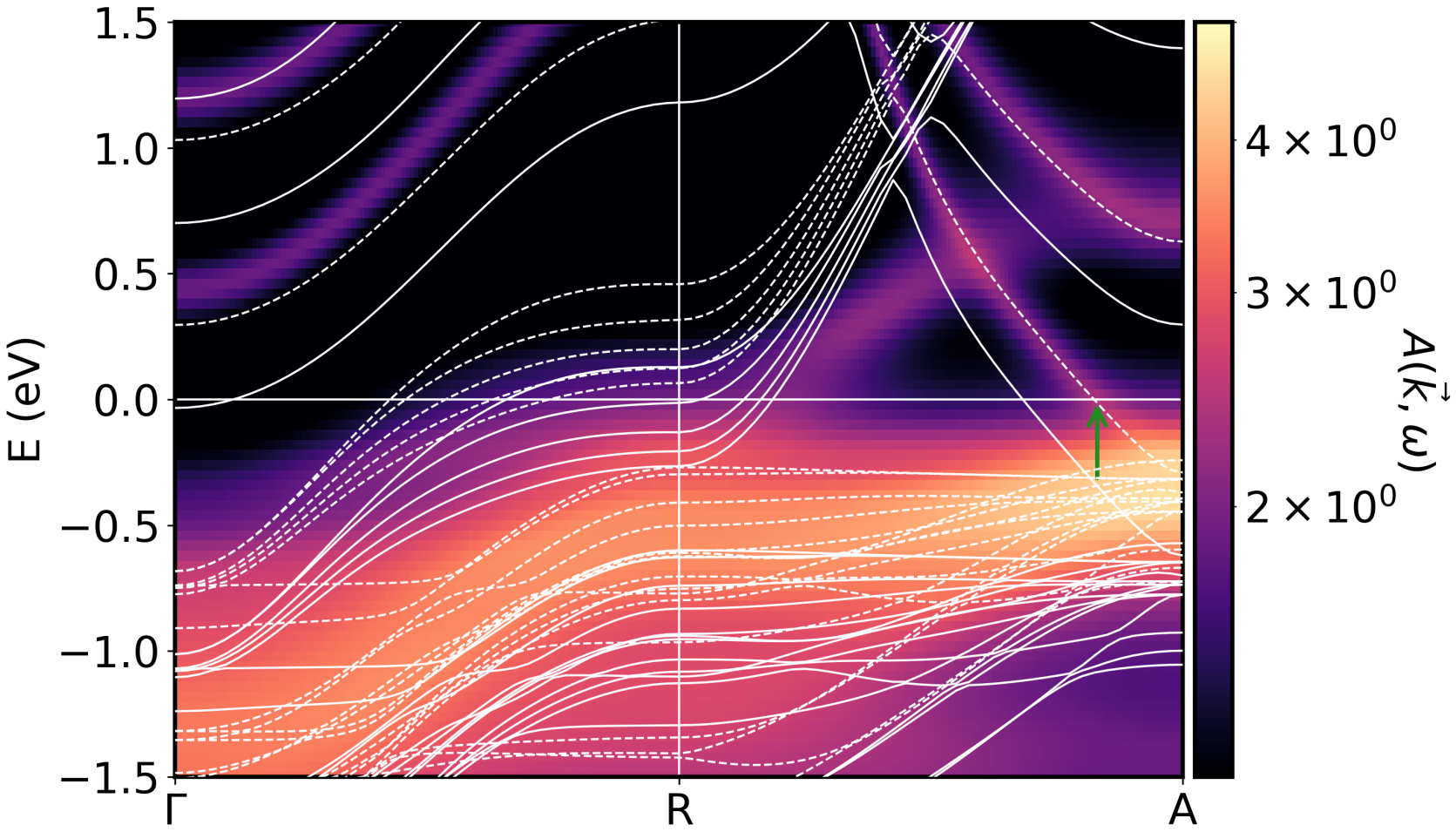

Supplementary Figure S12. Spectral function $A(\mathbf{k}, \omega)$ for $n = 5$ with DFT bands superimposed. The spectral function is indicated by the colormap, while unshifted DFT bands are shown as solid white lines, and shifted DFT bands as dashed white lines. The green arrow indicates the point in which the spectral function crosses the Fermi energy, which was used as reference to find the energy shift.

## SUPPLEMENTARY NOTE 6: EVALUATION OF $T_c$ WITH DΓA

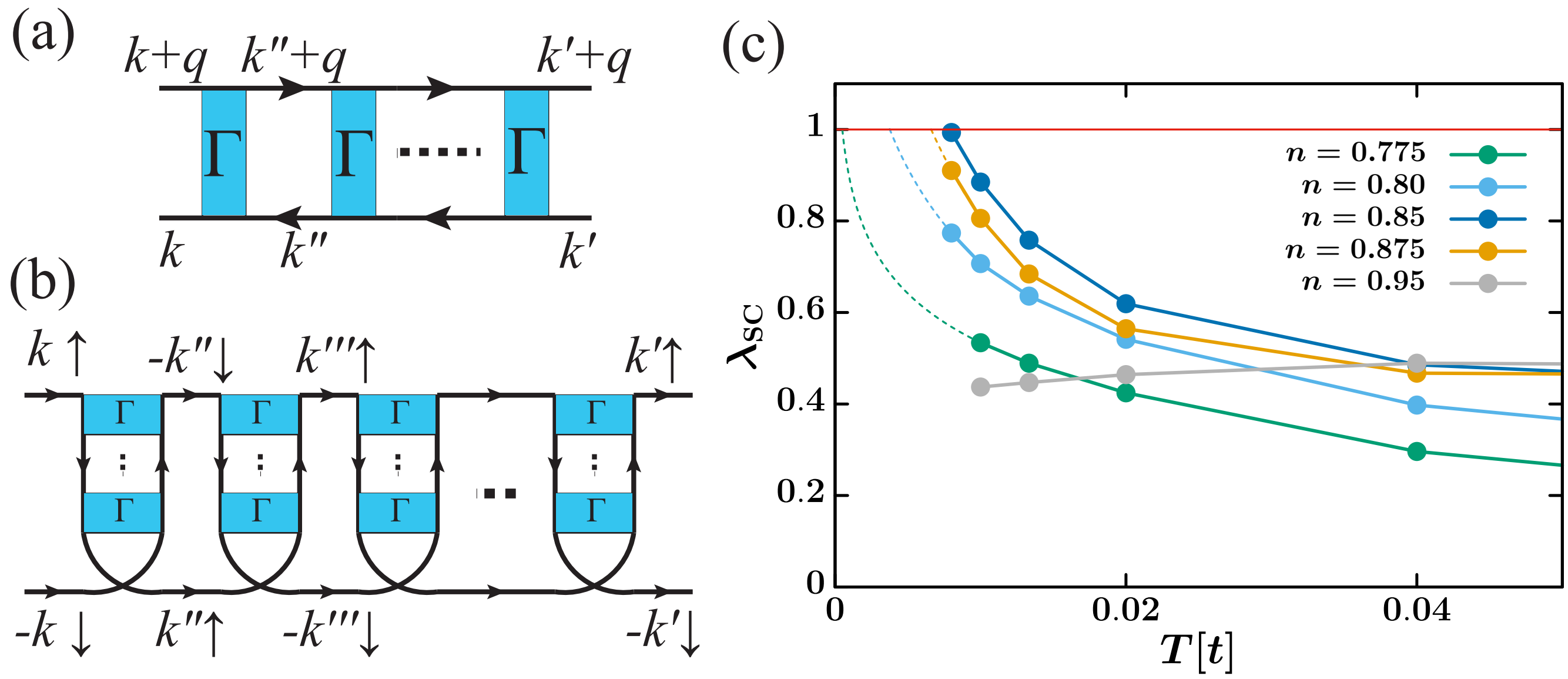

Supplementary Figure S13. (a) Diagrams represent DΓA spin fluctuations (solid line: the Green function, Γ: the irreducible vertex). (b) A particle-particle diagram represents superconducting instability mediated by DΓA spin fluctuations (an example diagram is shown). Adapted from [5, 6]. (c) The temperature dependence of the leading eigenvalue of the linearized gap equation in the square lattice Hubbard model with $t' = -0.25t, t'' = 0.12t$. Eigenvalue data taken from Ref. [7].

This section briefly reviews our formalism to determine $T_c$ with DΓA (cf., Refs. [5, 6]). DΓA is the typical method for diagrammatic extensions of the DMFT. In this approach, we incorporate the effect of non-local spin/charge fluctuations beyond DMFT [as depicted in Fig. S13(a)] into the self-energy. After obtaining the self-energy/Green function within DΓA we further consider the particle-particle fluctuations mediated by non-local particle-hole fluctuations, as illustrated in Fig. S13(b). To analyze the divergence of this particle-particle channel, we calculate the leading eigenvalue of the linearized gap equation as

$$\lambda\Delta(k) = \frac{-1}{\beta N}\sum_{k'}\Gamma_{\mathrm{pp}}(k,k',q=0)G(k')G(-k')\Delta(k'). \tag{S6}$$

Here, $G$ is the Green function, $N$ is the number of $k$-points, $\Gamma_{\mathrm{pp}}$ is the particle-particle irreducible vertex ($k, k', q$ are four-vectors consisting of Matsubara frequency and momentum, respectively), $\beta$ is the inverse temperature, and $\lambda$ ($\Delta$) is the eigenvalue (vector) of this equation. Fig. S13(c) shows the temperature dependence of the leading eigenvalue $\lambda$ of the current parameter set ($t' = -0.25t, t'' = 0.12t$). The transition temperature is determined where $\lambda$ reaches the unity. From these calculations, we can obtain the relation between the filling and the transition temperature (i.e., the phase diagram) within the single-band model.

As described in the main text, we evaluated the target single-orbital fillings in multilayer systems using three different methods: (i) the average DMFT $d_{x^2-y^2}$ occupation, (ii) the layer with optimal DMFT $d_{x^2-y^2}$ occupation, and (iii) $(1-1/n-\delta_{\mathrm{pocket}})/2$. Then, we determined the range of $T_c$ by comparing these fillings with the obtained phase diagram described above (The circles represent the midpoints of the determined $T_c$ ranges).

[4] To make sure the maximum entropy method for calculating the DMFT $A(\mathbf{k})$ is reliable, we have double checked it against a Fermi liquid fit $\mathrm{Re}\Sigma^{\mathrm{FL}}(\omega) = \mathrm{Re}\Sigma(\omega_1) + \frac{w}{w_1}\mathrm{Im}\Sigma(\omega_1)$ where $\omega_1$ is the lowest Matsubara frequency (not shown), which confirmed the maximum entropy analytical continuation.